\documentclass[aps,twocolumn,nofootinbib]{revtex4}
\usepackage{graphicx}
\usepackage{bm}

\begin{document} 

\title{\bf Metabolic evolution of a deep-branching hyperthermophilic chemoautotrophic bacterium}

\author{Rogier Braakman}

\affiliation{Santa Fe Institute, 1399 Hyde Park Road, Santa Fe, NM
87501, USA}

\author{Eric Smith}

\affiliation{Krasnow Institute for Advanced Study, George Mason
  University, 4400 University Drive, Fairfax, VA 22030, USA}

\affiliation{Santa Fe Institute, 1399 Hyde Park Road, Santa Fe, NM
87501, USA}

\date{\today}
\begin{abstract}

\emph{Aquifex aeolicus} is a deep-branching hyperthermophilic chemoautotrophic bacterium restricted to hydrothermal vents and hot springs. These characteristics make it an excellent model system for studying the early evolution of metabolism. Here we present the whole-genome metabolic network of this organism and examine in detail the driving forces that have shaped it. We make extensive use of phylometabolic analysis, a method we recently introduced that generates trees of metabolic phenotypes by integrating phylogenetic and metabolic constraints. We reconstruct the evolution of a range of metabolic sub-systems, including the reductive citric acid (rTCA) cycle, as well as the biosynthesis and functional roles of several amino acids and cofactors. We show that \emph{A. aeolicus} uses the reconstructed ancestral pathways within many of these sub-systems, and highlight how the evolutionary interconnections between sub-systems facilitated several key innovations. Our analyses further highlight three general classes of driving forces in metabolic evolution. One is the duplication and divergence of genes for enzymes as these progress from lower to higher substrate specificity, improving the kinetics of certain sub-systems. A second is the kinetic optimization of established pathways through fusion of enzymes, or their organization into larger complexes. The third is the minimization of the ATP unit cost to synthesize biomass, improving thermodynamic efficiency. Quantifying the distribution of these classes of innovations across metabolic sub-systems and across the tree of life will allow us to assess how a tradeoff between maximizing growth rate and growth efficiency has shaped the long-term metabolic evolution of the biosphere.

\end{abstract}

\maketitle

\section*{Introduction}

Metabolism lies at the heart of cellular physiology, acting as a chemical transformer between environmental inputs and components of biomass. Identifying rules and principles that underlie metabolic architecture can thus provide important insights into how basic properties of chemistry and physics constrain living systems. Of particular relevance to understanding the chemical history of the biosphere is the foundational layer of autotrophic metabolism, which fixes CO$_2$ and ultimately provides the ecological support to all forms of heterotrophy.

The merits of this view~\cite{Braakman13:Logic} were highlighted in a recent study on the early evolution of carbon-fixation pathways, which concluded that environmentally-driven innovations in this process underpin most of the deepest branches in the tree of life~\cite{Braakman12:C_fixation}. To extend our analysis of the early evolution of metabolism and of autotrophy, we present here a whole-genome reconstruction of the metabolic network of \emph{Aquifex aeolicus}. \emph{A. aeolicus} is a chemoautotroph, deriving both biomass and energy from inorganic chemical compounds, and is one of the deepest-branching and most thermophilic known bacteria~\cite{Huber06:Aquificales}. Deep-branching clades restricted to hydrothermal vents are generally considered to contain some of the most conservative metabolic features as a result of high degree of long-term stability provided by these environments~\cite{Reysenbach02:Vents}

While \emph{A. aeolicus} has been the focus of substantial experimental efforts (see Ref.~\cite{Guiral12:Aquifex_rev} for a review), it has not been characterized nearly as extensively as other model systems for which highly curated metabolic models exist.  In addition, the inherent uncertainty of genome annotation from sequence alone~\cite{Chothia86:Seq_fold,Schnoes09:annotation}, while overall significantly improving for next-generation methods~\cite{Radivojac13:Annotation}, is compounded by the deep-branching position and extremophile lifestyle of this organism. Metabolic reconstruction protocols generally rely on heuristic rules to deal with the inevitable network ``gaps'' that result from misannotation or the presence of genes of unknown function. Such protocols tend to perform well in predicting basic aspects of phenotype, such as growth rate, particularly for well-studied organisms~\cite{Feist09:networks,Henry09:B_subtilis}, but it is less clear what level of confidence to assign them when the focus is the evolution of specific metabolic sub-systems. Moreover, reconstructing an individual metabolic network requires substantial effort and provides only a single ``snapshot'' of an evolutionary process that has played out over several billion years.

For these reasons we utilize phylometabolic analysis (PMA)~\cite{Braakman12:C_fixation} to guide the reconstruction of the metabolic network of \emph{A. aeolicus} from its genome~\cite{Deckert98:Aquifex_genome}. PMA generates trees of functional metabolic networks (i.e. phenotypes) by integrating metabolic and phylogenetic reconstructions. The power of PMA derives
from a simple yet versatile constraint: the continuity of life in evolution.  Since metabolic pathways are the supply lines of monomers from which all life is constructed, the continuity of life requires
that at the ecosystem level \emph{some} pathway to a given universal metabolite \emph{must} be complete in any evolutionary sequence across different parts of the tree of life. The distribution of metabolic genes in different pathways to given metabolites, within and across clades, thus informs the most likely completions in individuals, while distributions of pathways suggest the evolutionary sequences that connect them (see also Methods). We recently introduced PMA to reconstruct the evolutionary history of carbon-fixation, relating all extant pathways to a single ancestral form~\cite{Braakman12:C_fixation}. Here we show the versatility of this approach, using it to reconstruct the complete whole-genome metabolic network of an individual species, while further examining
the evolutionary driving forces that have shaped the network.

As we will show, \emph{A. aeolicus} synthesizes a significant fraction of its biomass through metabolic pathways that appear to represent conserved forms of the ancestral pathways to those metabolites. This is relevant in debates on the position of this organism within the tree of life. Initial phylogenetic studies based on 16S rRNA suggested that the Aquificales represent potentially the deepest branch within the bacterial domain~\cite{Burgraff92:Aquif_phylo,Pace97:tree}. Later studies of conserved insertion-deletions (indels) in a range of proteins led to the conclusion that Aquificales are instead a later branch more closely related to $\epsilon$-proteobacteria~\cite{Griffiths04:Aquifex_late}, but this was subsequently found to be likely the result of substantial horizontal gene transfer (HGT) between these two clades~\cite{Boussau08:Aquifex_HGT}. It has since become clear that Aquificales and $\epsilon$-proteobacteria represent the dominant clades of primary producers near hydrothermal vents~\cite{Sievert12:vent_autotrophy}, and ecological association is now understood to be a major driver of HGT~\cite{Polz13:HGT_eco}. Together this appears to have restored some consensus on the very deep-branching position of \emph{A. aeolicus}~\cite{Boussau08:Aquifex_HGT,Zhaxybayeva09:thermotogales}, which is further supported by our analysis of its metabolism.

\subsection*{Innovations in metabolic evolution}

Our analysis highlights three classes of innovations in metabolic evolution. One of these is gene duplication and divergence along a progression from low substrate-specificity to high substrate-specificity enzymes, driven by selection for improved kinetics. It has long been argued that early enzymes had broader substrate affinities than modern enzymes, with greater potential to promiscuously catalyze homologous reactions in earlier times~\cite{Jensen76:recruitment,Obrien99:promiscuity,Copley03:promiscuity,Khersonsky10:promiscuity}. Broad affinity of ancestral enzymes is thought to have facilitated evolutionary adaptation by providing a `background' of biochemical pathways, initially proceeding at lower rates. If such background pathways produced advantageous products, they could then have been incorporated \emph{en bloc} into metabolism~\cite{Jensen76:recruitment}, initially through increased enzyme expression levels and eventually through duplication and divergence leading to more specific enzymes~\cite{Tawfik10:messy}. 

While selection for improved kinetics has probably lowered the overall occurrence of promiscuous enzymes, metabolism maintains a substantial degree of enzyme promiscuity. For example, \emph{E. coli} mutants in which an essential metabolic pathway was knocked out have been observed to recruit an alternate pathway from parts normally used for other functions to maintain growth~\cite{Kim10:serendipity}. In addition to promiscuity in the binding of alternate substrates while local functional group transformation is preserved (substrate promiscuity), promiscuity can also occur through the catalysis of alternate reaction mechanisms (catalytic promiscuity). The form of promiscuity most frequently encountered in extant cells appears to be substrate promiscuity~\cite{Khersonsky11:promiscuity_mode}. In general one might expect that the inherent ``messiness'' of enzymatic chemistry leads to a cost-benefit tradeoff in the evolution of substrate specificity, where complete specificity is difficult to achieve and moreover disadvantageous because it would decrease the capacity for future adaptation~\cite{Tawfik10:messy}.

In particular one would expect this tradeoff to be different for core processes, where a higher mass flux can significantly amplify the benefits of improved kinetics, versus more peripheral processes that have lower mass flux.  In keeping with this expectation, it is found that substrate promiscuity tends to increase toward the metabolic periphery~\cite{Nam12:network_specificity}, while reaction rate constants of enzymes tend to increase toward the metabolic core~\cite{Bar_Even11:enzyme_rates}.  The idea that selection for kinetics has determined the degree of specificity of a pathway's enzymes raises the intriguing possibility that prior to selection for increased substrate specificity, homologous reaction sequences could have initially been catalyzed by the same set of promiscuous enzymes~\cite{Jensen76:recruitment,Fondi07:LeuArgLys_evol,Pereto12:orig_metab}, allowing earlier metabolisms with greater abundances of such homologous sequences to be controlled with smaller genomes. 

We will discuss the evolution of several sub-networks in the metabolism of \emph{A. aeolicus} that provide illustrations of these general principles. We will show that compared with later branching autotrophs \emph{A. aeolicus} uses a greater abundance of repeated parallel chemical sequences catalyzed by enzymes with high sequence similarity, which could have initially been catalyzed by a smaller number of more promiscuous enzymes. We will further highlight how cost-benefit tradeoffs of improving kinetics have played out differently in core versus peripheral pathways.

A second class of innovations concerns the fusion of enzyme subunits into larger single-subunit enzymes. Enzyme fusion increases the effective concentrations of reactants at the active site for subsequent reactions within a sequence, thus increasing the throughput rate of the sequence~\cite{Marcotte99:fusion}. In particular when intermediates within pathway sequences are not used elsewhere in metabolism, the fusion of associated enzymes would appear to have potentially little cost, but significant kinetic benefits for pathways with high mass-flux densities. Accordingly, studies have generally shown that gene fusion events significantly outnumber gene fission events in evolution~\cite{Snel00:fusion,Kummerfeld05:fusion}. Similarly, the organization of enzymes into multi-subunit enzymes, or even super-complexes, can, in addition to facilitating novel reaction mechanisms, also be considered to improve the kinetics of metabolism by increasing effective concentrations of intermediates. While this is not a central focus of the current study, we will highlight several key reaction sequences that in \emph{A. aeolicus} are catalyzed by multi-subunit enzymes that in later branching organisms are known to be fused, reflecting \emph{Aquifex}'s more primitive metabolism.

A final important class of innovations occurs at the level of pathways.  In some cases organisms may have access to multiple different pathways which produce certain metabolites, with different pathways' having different unit costs of required ATP hydrolysis. Recent work has suggested that lowering the ATP cost, thereby increasing overall thermodynamic efficiency, of pathways involved in CO$_2$-fixation was a major evolutionary driving force that resulted in several major early branches in the tree of life~\cite{Braakman12:C_fixation}. Here we show additional divergences that appear to be driven by increasing thermodynamic efficiency.

These three classes of innovations are all different expressions, at different levels, of a more general evolutionary tradeoff, in which either the efficiency or the rate of growth is maximized. For example, for heterotrophic organisms it is observed that in the presence of resource competition cells will use fermentative metabolic modes and maximize the rate of ATP production, while in the absence of competition cells will use respiration to maximize the yield of ATP production~\cite{Pfeiffer01:RK_selec}. Because ATP hydrolysis drives biosynthesis, increasing its production rate will increase growth rate, while increasing ATP yield will increase growth efficiency. 

A second aspect of metabolism where this growth rate/efficiency tradeoff is expressed, is not in the production of ATP, but in the structure of the biosynthetic machinery driven by its subsequent hydrolysis. For example, improving kinetics of metabolic sequences through increased substrate specificity or gene fusion contributes to improving growth rate, while lowering ATP cost by choosing alternative pathways contributes to improving efficiency. The relative mass contributions of different pathways or reactions are likely a critical determinant in the balance of benefit vs. cost of such innovations in different parts of metabolism.

Cost-benefit tradeoffs in the structure of biosynthesis are likely to be of particular importance in autotrophic metabolism. Because (some) heterotrophs can switch between fermentative and respiratory strategies for given organic inputs, they have access to much larger variability in the rate of ATP production than do autotrophs. Autotrophs generally use purely respiratory metabolic modes to interconvert pairs of inorganic compounds and are thus more constrained on the ATP production side. Since autotrophs form the metabolic basis for ecosystems, the same will tend to hold for aggregate metabolic networks at that level. The biosphere is globally autotrophic, and especially on the longest time-scales of selection we may thus expect that just as new inorganic energy sources were being explored to increase total inputs to ecosystems, both the rate and efficiency of metabolic processes were simultaneously being optimized where possible. Even if individuals have access to, and can switch between alternate strategies and lifestyles, both those individuals as well as the ecosystem to which they belong would benefit from improved kinetics and efficiency of their metabolisms.  The present reconstruction of the metabolism of \emph{A. aeolicus} provides an excellent testbed to start cataloging and exploring these ideas.

\section*{Materials and Methods}

The basic principles of phylometabolic analysis (PMA)~\cite{Braakman12:C_fixation} are outlined if Fig.~\ref{PMA}. The effectiveness and flexibility of PMA for studying metabolic evolution arises directly from the strongly synergistic potential of integrating metabolic and phylogenetic constraints into a single approach. For example, if an individual annotated genome leads to a putative metabolic network that is not viable due to gaps in all known pathways that connect given sub-networks, then without experimental evidence it may be very unclear how to proceed in the curation process. As shown in panel A, placing that same putative network in phylogenetic context and comparing metabolic gene profiles within and across clades, often clearly suggests the proper gap-fills. We will show many such examples of the ways PMA guides the reconstruction process for the metabolic network of \emph{A. aeolicus}.

The same genomic surveys that help interpret the character of individual phenotypes lead also to a distribution of metabolic pathways across the tree of life, as shown in panel B. PMA exploits the existence of universal topological bottlenecks in metabolism -- metabolites through which all pathway alternatives in a given sub-system must pass to allow biosynthesis. Imposing a requirement on reconstructions that all extant and ancestral metabolic states supply those bottleneck metabolites then allows us to transform the distributions of pathways into specific evolutionary sequences. These sequences can be represented as phylometabolic trees, as shown on the right in panel B. 

In comparison to phylogenetics, which reconstructs evolutionary trees through analyzing presence/absence of genes or the sequence similarity of genes, PMA thus generates trees by analyzing the presence/absence of pathways or the similarity of (sub)networks.  Moreover, because each node in a phylometabolic tree represents a distinct functional phenotype with an explicit internal chemical structure, the comparison between nodes can often identify clear physical-chemical driving forces underlying divergences~\cite{Braakman12:C_fixation}. For \emph{A. aeolicus} this helps us identify the key evolutionary driving forces that have shaped its metabolism.

\begin{widetext}
\begin{center}
\begin{figure}[h]
\begin{center}
\includegraphics[scale=0.9]{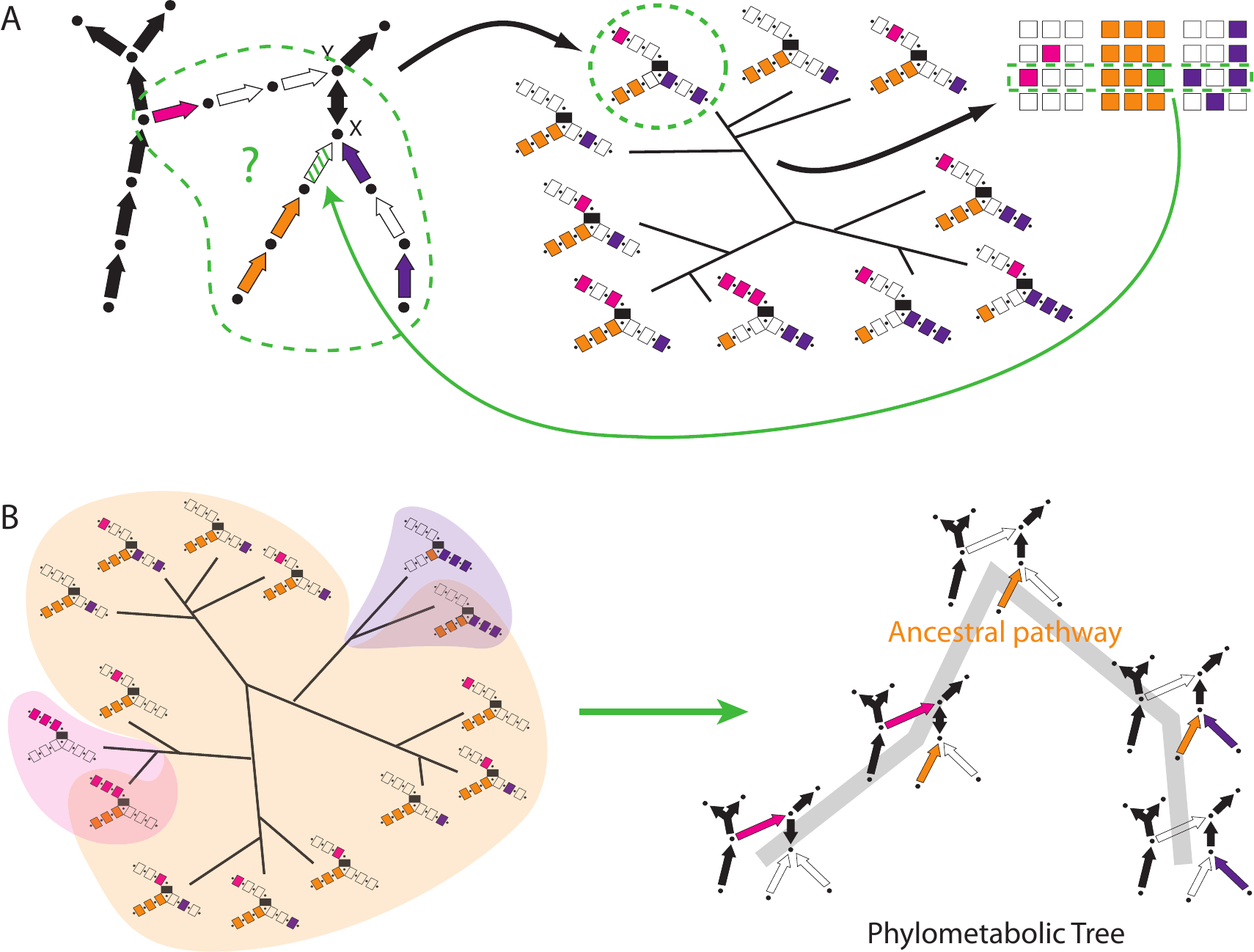}
\end{center}
\caption{Principles of Phylometabolic Analysis. Panel A shows how phylogenetic distributions of pathways helps interpret the curation of individual metabolic networks. In this case comparison of metabolic gene profiles suggests the orange pathway represents the correct completion. Panel B in turn shows how pathway distributions in turn also suggest evolutionary sequences. Imposing continuity of metabolite production at the ecosystems level allows us to represent those sequences as phylometabolic trees, in which each node represents a functional phenotype with an explicit internal chemical structure.}
\label{PMA}
\end{figure}
\end{center}
\end{widetext}

A critical aspect of PMA is that the evolutionary sequences represented by phylometabolic trees are best understood at the ecosystem level. Enforcing continuity in the overall connectedness of metabolism requires the existence of evolutionary intermediates in which two pathways are co-present before one may replace another (see Fig.~\ref{PMA}). However, such intermediary states are not required to exist within a single organism, as an individual may give up the ability to use one pathway before evolving or incorporating (through gene transfer) a second to replace it, so long as in the interim it can take up the required bottleneck intermediate from an ecosystem that continues to produce it.  Moreover, since in PMA we generally model metabolic sub-systems without explicitly considering aspects of regulation or
cellular integration, the approach itself also does not distinguish between individuals and ecosystems.  The nodes in a reconstructed tree thus capture the sequence of dependencies in the innovation of pathways, whether these occurred within single species or in syntrophic consortia. Nevertheless, the metabolism of individual species may still contain most or all of the network contained in individual nodes of a phylometabolic tree. This is particularly true for autotrophs (such as \emph{A. aeolicus}), which generate all biomass components directly
from inorganics and are in a sense ``ecosystems within individuals'', and thus form important model systems for reconstructing the long-term evolution of metabolism.

Our reconstruction of a whole-metabolism model for \emph{A.  aeolicus} is based on an initial network obtained from the Model SEED server~\cite{Henry10:SEED}, an automated pipeline that generates genome-scale metabolic models directly from genome annotations. After first modifying the nutrient and biomass compositions of the model to accurately capture the boundary conditions that define the overall \emph{A. aeolicus} phenotype (see supplemental info for additional details), the internal network was curated, using criteria of phylometabolic consistency to evaluate proposed completions of of key network gaps.

In practice, PMA was implemented by surveying the annotated genomes of a large number of archaea and bacteria across the tree of life at the online Uniprot database~\cite{Uniprot:uniprot_update:11}. Other repositories may be used, but we chose Uniprot because it derives from the manually curated Swissprot database, which is known to have among the lowest error rates in gene annotation~\cite{Schnoes09:annotation}. We then performed an exhaustive and redundant search for all relevant genes, using naming variants and EC classification numbers, which code for the enzymes across a set of metabolic pathways that define a metabolic sub-system.  When all enzymes of a metabolic pathway are identified within a single genome, then that pathway is counted as present in that species, and totals are tabulated across clades. The high rate at which new genomes are being sequenced is reflected in the fact that the number of members in clades differs in our tabulations for different sub-systems, even though analyses were in some cases only performed a few weeks apart.

To further correct for misannotated or unannotated genes, the above searches were complemented by additional BLAST searches using the built-in capabilities on the Uniprot website. As we will highlight with several examples, in many cases BLAST searches will identify groups of genes coding for the same enzyme at the clade level. Especially when organisms contain closely related enzymes that because of high sequence similarity are misannotated (often as the same gene), comparisons using both sequence similarity and enzyme length will often separate those enzymes into clear groups at the clade level. This in turn increases the chances that individual enzymes within those clade-level groups have been experimentally studied within an individual member of that clade, helping to anchor the functional annotation of each of the enzyme groups.

\section*{Results and Discussion}

\emph{Aquifex aeolicus} provides an interesting reference point for minimal deep-branching hyperthermophilic bacterial metabolism. Our reconstruction for this organism contains 756 reactions and 729 metabolites. In comparison, the recently reconstructed network of \emph{Thermotoga maritima} -- like \emph{A. aeolicus} thought to be one of the deepest-branching and most thermophilic bacteria~\cite{Zhaxybayeva09:thermotogales} -- contains 562 reactions and 503 metabolites~\cite{Zhang09:thermotoga}. Most of the difference in size results from the ways fatty acid metabolism is represented. In the \emph{T.  maritima} network, lipid metabolism is represented using many composite reactions, resulting in only 27 reactions and 27 metabolites in this sub-network~\cite{Zhang09:thermotoga}. By contrast, we explicitly represent each reaction in lipid synthesis, and because the lipid composition of \emph{A. aeolicus} is diverse, this results in nearly 200 reactions and 200 metabolites (see also the section on lipid biosynthesis below). However, fatty acid biosynthetic machinery is highly flexible and modular, with chains of diverse length and substitution patterns generated by a very small set of proteins~\cite{White05:FA_biosynth}. It should further be noted that \emph{T. maritima} in fact also has a diverse mixture of chain length and substitution patterns in its lipid content~\cite{Carballeira97:FA_hyperthermo}, but that the simpler composition of \emph{E. coli} was chosen in defining its biomass for reconstruction~\cite{Zhang09:thermotoga}. Thus, taking this representation difference into account, the metabolic networks of \emph{A. aeolicus} and \emph{T. maritima} may be reasonably representative of minimal bacterial metabolism found near hydrothermal vents.

Our primary goal was to accurately reconstruct the pathway structure of the metabolism for the purpose of understanding its evolution. We therefore did not significantly modify the biomass vector of the model (with fatty acids as the major exception), or general assignments of genes to enzymes from the initial model obtained from the SEED server. This work should thus be considered a qualitative first step toward building a comprehensive computational organism model for \emph{A. aeolicus}. As additional experimental data become available, including in particular detailed data on the biomass composition under different growth conditions, they will provide further benchmarks to make this model increasingly quantitative. Nevertheless, this reconstruction contains a wealth of data on basic principles of metabolic evolution, as we will discuss in detail, starting from core carbon-fixation.

\subsection*{Carbon-fixation and the initiation of anabolism}

\emph{Aquifex aeolicus} uses a unique and previously unrecognized form of carbon-fixation. It has been known for more than two decades that \emph{A.  aeolicus} uses the reductive citric acid (rTCA) cycle to fix most of its carbon~\cite{Beh93:rTCA,Hugler07:Aquifex_2TCAs}. However, our study on the evolution of carbon-fixation led us to conclude that in parallel this organism uses an incomplete form of the reductive acetyl-CoA (Wood-Ljungdahl, WL) pathway to produce a small sub-set of biomass components~\cite{Braakman12:C_fixation}. This hybrid form of carbon-fixation was recognized through a broad survey of the biosynthetic pathways leading to glycine and serine. We identified substantial gaps in each of the conventionally recognized pathways to these compounds across many deep-branching clades. Instead these organisms often possess the complete gene complement for an alternate folate based reductive C$_1$ pathway that is also used as part of WL. Indeed, the suggested existence of widespread hybrid carbon-fixation strategies using a partial WL pathway was one of the key insights that allowed the reconstruction of a phylometabolic tree of carbon-fixation through fully autotrophic intermediates~\cite{Braakman12:C_fixation,Braakman13:Logic}.

\begin{figure}[!ht]
\begin{center}
\includegraphics[scale=0.48]{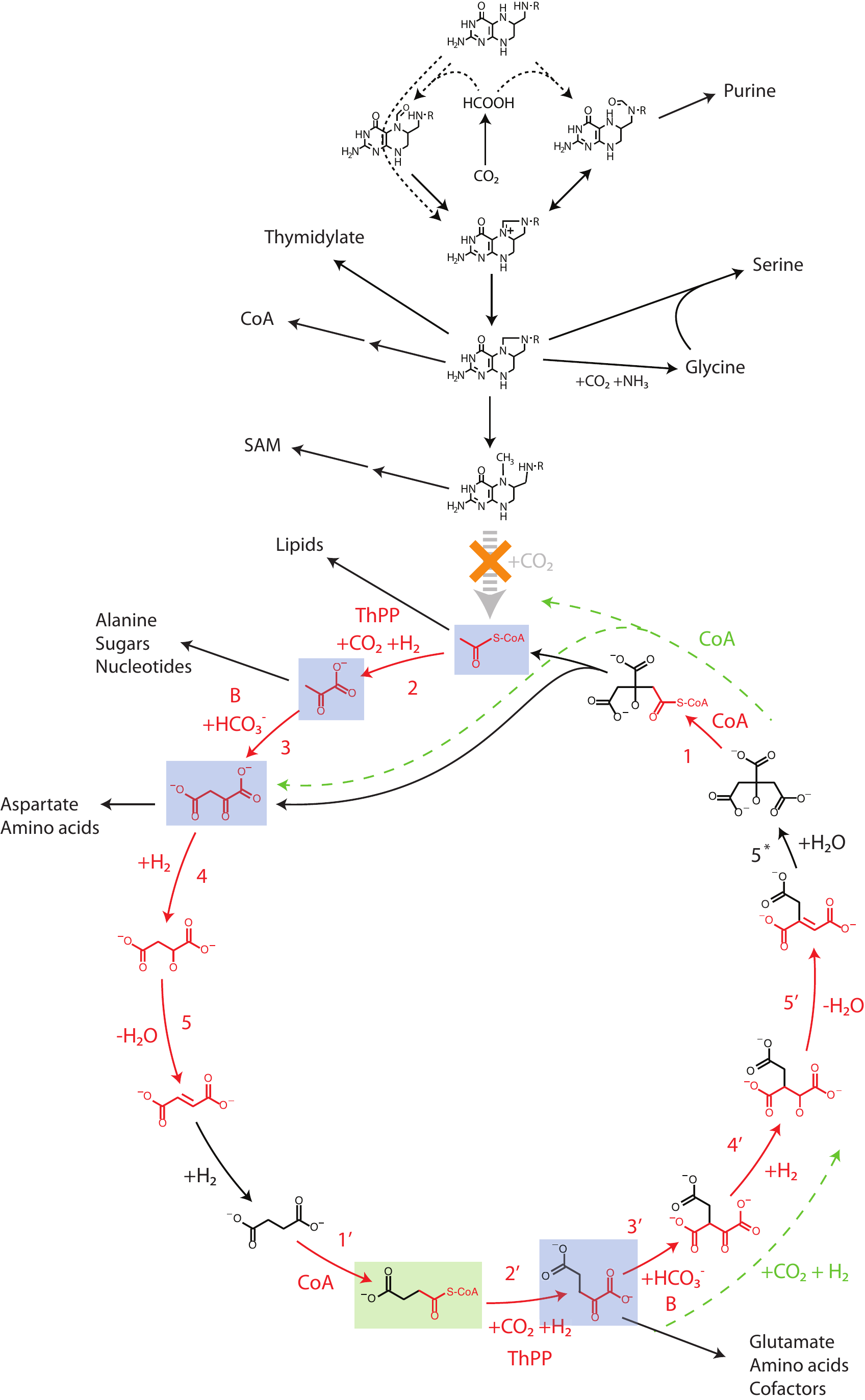}
\end{center}
\caption{Carbon-fixation in \emph{Aquifex aeolicus}. The main fixation pathway is the reductive citric acid (rTCA) cycle, from which most anabolic pathways are initiated. Reductive folate chemistry is a secondary fixation pathway from which an additional small set of anabolic pathways is initiated. Relative to the reconstructed root of carbon-fixation, in which the rTCA cycle and Wood-Ljungdahl pathway are fully integrated~\cite{Braakman12:C_fixation}, this hybrid strategy employed by \emph{A. aeolicus} lacks only the grey-dashed reaction (acetyl-CoA synthesis). Molecules highlighted in blue represent the ``pillars of anabolism'', TCA intermediates from which the vast majority of anabolic pathways have been initiated throughout evolution~\cite{Smith04:universality}. Highlighted in green is succinyl-CoA, which forms a precursor to pyrroles through a later derived pathway in some organisms (but not Aquifex). Highlighted in red are reaction sequences involving the same local functional group transformation that in \emph{A. aeolicus} are catalyzed by closely related enzymes in both halves of the rTCA cycle. Green dashed arrows highlight alternate pathway sequences catalyzed by a single enzyme in other clades.}
\label{fig:C_Fix}
\end{figure}

The complete carbon-fixation strategy reconstructed for \emph{A. aeolicus} is shown in Fig.~\ref{fig:C_Fix}. The major fraction of carbon is fixed through the rTCA cycle, producing its intermediates acetyl-CoA, pyruvate, oxaloacetate, and $\alpha$-ketoglutarate (highlighted in blue), from which anabolic pathways to most components of biomass subsequently radiate~\cite{Smith04:universality}. In parallel, reductive folate chemistry transforms CO$_2$ into C$_1$ units of different oxidation states, from which a small number of additional anabolic pathways are initiated. Formyl groups are used in the production of purine, methylene groups are used in the production of glycine and serine, as well at thymidylate and coenzyme A, while methyl groups are donated to S-adenosyl methione (SAM), which mediates methyl-group chemistry~\cite{Maden00:folates}. This combined carbon-fixation strategy thus lacks only a single reaction relative to the reconstructed root of carbon-fixation, in which the rTCA cycle and the WL pathway are fully integrated~\cite{Braakman12:C_fixation}. The \emph{A. aeolicus} metabolic phenotype is lacking only the final synthesis of acetyl-CoA within WL, which is catalyzed by one of the most oxygen-sensitive enzymes in the biosphere~\cite{Ragsdale83:CODH}.

The specific form of formate uptake by folate remains to be elucidated. \emph{A. aeolicus} lacks the gene for 10-formyl-THF (tetrahydrofolate) synthetase, which catalyzes the attachment of formate at the N$^{10}$ position of THF in the acetogenic version of WL~\cite{Maden00:folates}. However, a broad genomic survey across deep branches of the tree of life nonetheless strongly supports the use of a partial WL pathway over other alternatives to produce glycine and serine in \emph{A. aeolicus}~\cite{Braakman12:C_fixation}.

Formate incorporation in this organism is most likely catalyzed by an unrecognized N$^{10}$-uptake protein, or through an unrecognized sequence involving attachment of formate at the N$^5$ position of THF~\cite{Braakman12:C_fixation}. This alternate route was suggested by the very wide distribution across the tree of life of a gene for an ATP-dependent 5-formyl-THF cycloligase, present in the genomes of many organisms that like \emph{A.  aeolicus} appear to use a partial WL pathway to produce glycine and serine, but which lack the (also ATP-dependent) 10-formyl-THF synthetase. 5-formyl-THF cycloligase catalyzes the cyclization of 5-formyl-THF to methenyl-THF and has so far only been suggested to be part of futile cycle, but its precise physiological role has long remained unclear~\cite{Stover93:5FTHF_cyclo,Huang95:5FTHF_cyclo}. The suggested use of an N$^5$ uptake route for formate is appealing not only for these reasons, but also because it would provide an evolutionary intermediate between N$^5$ formate uptake in the methanogenic version of WL and N$^{10}$ formate uptake in the acetogenic version of WL.

While it has been recognized that most anabolic pathways originate from TCA intermediates, there exist some variation in this set of intermediates across clades. In particular, many organisms derive pyrroles (precursors to the cofactor family that includes heme and chlorophyll) from $\alpha$-ketoglutarate through what is known as the C5 pathway, while $\alpha$-proteobacteria and mitochondria derive pyrroles from succinyl-CoA~\cite{Wettstein95:chlorophyll}. This had previously already led to the suggestion that pyrrole synthesis from succinyl-CoA is a later evolutionary innovation~\cite{Benner89:palimpsest}, which is confirmed by genomic surveys. Table~\ref{tab:pyrroles} shows that the pathway from succinyl-CoA is almost completely absent from deep-branching bacterial and archaeal clades, while the C5 pathway is nearly universally distributed. Thus, across most of the tree of life, and for most of metabolic evolution, carbon-fixation has fed into anabolism through only 4 TCA intermediates.

\begin{table}[ht]
\footnotesize
\begin{tabular}{|l|ll|}
\hline
  &  C5 & Aminolevulinate \\
  Domain & & \\
\hline
 Archaea (133)  & 108 & 0 \\ 
 Bacteria$^{*}$ (345) & 290 & 5\\
 \hline
\end{tabular}
\caption{
\footnotesize Distribution of entry sequences to pyrrole biosynthesis. Notes: $^{*}$ Includes Aquificales, Thermotogales, Firmicutes (Bacillales + Clostridia), Chloroflexi, Chlorobiales, Cyanobacteria, Nitrospirae, Deincoccales, Verrucomicrobia, Planctomycetes.
\label{tab:pyrroles}
}
\end{table}

\subsubsection*{Evolution of the rTCA cycle}

Variants in the form of the rTCA cycle used within the Aquificales provide insights into the evolutionary driving forces that have shaped this pathway. Recent studies in \emph{Hydrogenobacter thermophilus}, which together with \emph{A. aeolicus} belongs to the Aquificaceae family within Aquificales, identified novel enzymes for several rTCA reactions. The cleavage of citrate to acetyl-CoA and oxaloacetate, and the reductive carboxylation of $\alpha$-ketoglutarate to isocitrate, each conventionally recognized as single enzyme reactions, are in \emph{H. thermophilus} both catalyzed in two steps by distinct enzymes also found in the \emph{A. aeolicus} genome (see Fig.~\ref{fig:C_Fix})~\cite{Aoshima04:CIT_synthase,Aoshima04:CIT_lyase,Aoshima04:AKG_carbox,Aoshima06:Oxalosucc}. This Aquificaceae variant of the rTCA cycle has an increased degree of symmetry, and reveals previously unrecognized homology relations in enzymes, which recapitulate the similarity in the local-group chemistry of their substrates.

The newly discovered citryl-CoA synthase and $\alpha$-ketoglutarate biotin carboxylase enzymes catalyze local functional group chemistry that is homologous to their counterparts succinyl-CoA synthase and pyruvate biotin carboxylase (reactions 1, 1$^{\prime}$ and 3, 3$^{\prime}$ in Fig.~\ref{fig:C_Fix}), respectively. Moreover, both sets of homologous enzymes have high sequence similarity, and were originally annotated as the same enzymes~\cite{Aoshima04:CIT_synthase,Aoshima04:AKG_carbox}. Similarly, pyruvate and $\alpha$-ketoglutarate ferredoxin:oxidoreductase also catalyze homologous chemistry (reactions 2, 2$^{\prime}$), and due to high sequence similarity were again annotated as the same enzyme in \emph{A. aeolicus}~\cite{Deckert98:Aquifex_genome}.

These observations are striking in light of the discussion on metabolic innovations in the introduction. A hypothesis of a minimal ancestral metabolism with promiscuous enzymes catalyzing homologous reactions in parallel, is consistent with a highly symmetric rTCA variant used by members of the Aquificaceae. Aquificales are the deepest-branching clade using this pathway to fix carbon~\cite{Hugler11:C_fix}, and as mentioned their exclusive association with hydrothermal vents may result in some of the most conservative metabolic features~\cite{Reysenbach02:Vents}.

Indeed, the new enzymes identified in the Aquificaceae have been argued to represent the ancestral rTCA enzymes~\cite{Aoshima04:CIT_lyase,Hugler07:Aquifex_2TCAs,Aoshima08:ICIT_dehydro}. The single step ATP citrate lyase found in most rTCA bacteria is suggested to have arisen through a gene fusion of the second sub-unit of citryl-CoA synthase with citryl-CoA lyase~\cite{Aoshima04:CIT_synthase,Aoshima04:CIT_lyase,Hugler07:Aquifex_2TCAs}, while the combination of $\alpha$-ketoglutarate biotin carboxylase plus isocitrate dehydrogenase (ICDH) is suggested to have been replaced by a single ICDH with increased substrate specificity~\cite{Aoshima08:ICIT_dehydro}.

To further explore the evolutionary driving forces that underly the evolution of the rTCA cycle, we used \emph{H. thermophilus} enzymes as a benchmark to survey the genomes of all Aquificales, as well as other clades using the rTCA cycle. Table~\ref{tab:rTCA_genes} shows the distributions of each of the newly identified high homology rTCA enzymes across all Aquificale genomes.

\begin{widetext}
\begin{center}
\begin{table}[ht]
\footnotesize
\begin{tabular}{|ll|l|llll|llll|}
\hline
  && \multicolumn{1}{c|}{ICDH} & \multicolumn{2}{c}{BC - LS} & \multicolumn{2}{c|}{BC - SS} & \multicolumn{2}{c}{CS - LS} & \multicolumn{2}{c|}{CS - SS} \\
  Family &Species & & PYR~~~~ & AKG~~~~~~~~ & PYR~~~~ & AKG~~~~~~~~ & SUC-~~~~ & CIT-~~~~~~~~ & SUC-~~~~ & CIT-~~~~~~~~\\
\hline
 I & \emph{T. thermophilus} & ~~~~421~~~ & 617 (100) & 652 (100) & 475 (100) & 472 (100) & 383 (100) & 429 (100) & 293 (100) & 344 (100) \\
 & \emph{A. aeolicus} & ~~~~426~~~ & 614 (81) & 655 (78) & 477 (76) & 472 (82) & 385 (77) & 436 (78) & 305 (84) & 368 (84) \\
 & \emph{T. Albus} & ~~~~419~~~ & 614 (81) & 653 (90) & 475 (83) & 472 (91) & 382 (81) & 432 (87) & 295 (87) & 341 (89) \\
 & \emph{H. sp Y04AAS1} & ~~~~421~~~ & 619 (69) & 638 (74) & 475 (69) & 472 (79) & 382 (74) & 423 (78) & 291 (80) & 329 (77) \\
\cline{1-11}
 II & \emph{S. azorense} & ~~~~421~~~ & 614 (65) & 647 (74) & 482 (67) & 472 (73) & 388 (63) & $\ast\ast$~~~(31) & 293 (72)& $\ast\ast$~~~(39) \\
 & \emph{P. marinus} & ~~~~747~~~ & 613 (67) & $\ast\ast$~~~(50) & 478 (64) & $\ast\ast$~~~(56) & 388 (65) & $\ast\ast$~~~(31) & 292 (69) & $\ast\ast$~~~(38) \\
 & \emph{S. sp YO3AOP1} & ~~~~746~~~ & 616 (66) & $\ast\ast$~~~(47) & 476 (66) & $\ast\ast$~~~(56) & 389 (64) & $\ast\ast$~~~(30) & 293 (72) & $\ast\ast$~~~(39) \\
\cline{1-11}
III & \emph{D. thermolithotrophum} & ~~~~735~~~ & 616 (57) & $\ast\ast$~~~(50) & 472 (58) & $\ast\ast$~~~(54) & 389 (62) & $\ast\ast$~~~(31) & 300 (70) & $\ast\ast$~~~(37) \\ 
 & \emph{T. ammonificans} & ~~~~735~~~ & 618 (59) & $\ast\ast$~~~(49) & 472 (59) & $\ast\ast$~~~(54) & 389 (63) & $\ast\ast$~~~(30) & 300 (70) & $\ast\ast$~~~(37) \\
\hline
\end{tabular}
\caption{
Distribution of rTCA enzymes in Aquificales. Numbers in each column represent the length of the enzyme in terms of amino acid residues, and the numbers in parentheses are the sequence similarities relative to the \emph{T. thermophilus} version of those enzymes. Entries $\ast\ast$ are for species that have only one copy of that enzyme class, which are aligned with the type to which it has greatest sequence similarity (sequence similarity for alternate alignment also shown for comparison). Families: I - Aquificaceae, II - Hydrogenothermaceae, III - Desulfurothermaceae. Abbreviations: ICDH, isocitrate dehydrogenase; BC, biotin carboxylase; -CoA synthase; LS, large subunit; SS, small subunit; PYR, pyruvate; AKG, $\alpha$-ketoglutarate; SUC-, succinyl-; CIT-, citryl.
\label{tab:rTCA_genes}
}
\end{table}
\end{center}
\end{widetext}

Two clear groups of Aquificales are distinguished by the rTCA variants they use. All members of the Aquificaceae possess two biotin carboxylase (BC) and two CoA synthase (CS) enzymes. By contrast, all member of the Hydrogenothermaceae and Desulfurobacteriaceae possess only single copies of these enzyme types, except \emph{S. azorense}, which possess two BC enzymes. In addition, in terms of both sequence similarity and length, both sub-units of the single BC enzyme in Hydrogenothermaceae and Desulfurobacteriaceae best match pyruvate (rather than $\alpha$-ketoglutarate) BC in Aquificaceae. Similarly, the single CS enzyme in these families best matches succinyl-CoA synthase in Aquificaceae. Finally, all Hydrogenothermaceae and Desulfurobacteriaceae are known to possess ATP citrate lyase, while all except \emph{S.  azorense} possess an ICDH enzyme that is significantly larger than the version found in Aquificaceae. This difference in length is consistent with the suggestion that an increased substrate specificity of the ICDH allowed it to supplant the combined function of the BC and more primitive ICDH enzymes~\cite{Aoshima08:ICIT_dehydro}. This survey shows how sequence length of enzymes can be a useful secondary source of information, after sequence similarity, in functional annotation. Thus, we conclude that all Aquificaceae possess the highly symmetric rTCA variant shown in Fig.~\ref{fig:C_Fix}, while other Aquificale families mostly possess the more conventional rTCA variant.

Clear evolutionary driving forces can be identified that connect the different extant forms of the rTCA cycle.  The gene fusion within ATP citrate lyase increases the effective concentration of citryl-CoA in the subsequent cleavage reaction, and can thus be understood to improve the kinetics of the rTCA cycle. This gene fusion would appear to have low cost, not requiring evolution of an additional compensatory pathway to release citryl-CoA as a free intermediate, since it is not used elsewhere in metabolism.

Detailed thermodynamic analyses have in turn identified two classes of rTCA reactions that in isolation would require ATP hydrolysis to proceed: carboxylation reactions and carboxyl reduction reactions~\cite{BarEven12:CFix_thermo}. These costs can be avoided in a number of ways. Coupling carboxyl reductions to subsequent carboxylations through thioester intermediates (e.g.  succinyl- or acetyl-CoA), allows the combined sequence to be driven by a single ATP hydrolysis, while coupling endergonic to subsequent exergonic reactions can eliminate the ATP cost of the combined sequence~\cite{BarEven12:CFix_thermo}. The replacement of the combination of $\alpha$-ketoglutarate BC and the lower-specificity ICDH with a single high-specificity ICDH falls into this latter category. The $\alpha$-ketoglutarate BC reaction in Aquificaceae costs 1 ATP per carbon incorporated~\cite{Aoshima04:AKG_carbox}, while the subsequent reduction to isocitrate is highly exergonic, resulting in a nearly-reversible sequence without ATP cost when the reactions are coupled~\cite{Miller:TCA_Gibbs:90,Smith04:universality,BarEven12:CFix_thermo}. This adaptation can thus be seen as driven by improving the thermodynamic efficiency of the rTCA cycle.

An additional factor that could have further increased the ATP savings of using a single higher-specificity ICDH to generate isocitrate from $\alpha$-ketoglutarate is that like other $\beta$-ketoacids oxalosuccinate is subject to spontaneous decarboxylation. The lifetime of oxaloacetate (also a $\beta$-ketoacid) in water may be estimated at about 3~min at ${90}^{\circ} \, {\rm C}$ (Figure based on $\Delta H^{\ddagger} = 17.2 \, \mbox{kcal/mol}$ and $k_{{25}^{\circ} {\rm C}} \approx 2.8 \times {10}^{-5} \, {\rm s}^{-1}$ from Ref.~\cite{Wolfenden:decays:11}.), and oxalosuccinate is more unstable. Metal ions and amines are known, at least for oxaloacetate, to further enhance the rate of spontaneous decarboxylation~\cite{Wolfenden:decays:11}. The decarboxylation of oxalosuccinate back to its precursor $\alpha$-ketoglutarate would introduce a futile cycle that may have raised the average cost of the forward carboxylation reaction above 1 ATP.

Surveys in the genomes of other clades using the rTCA cycle (data not shown) are consistent with these suggested adaptations. Chlorobiales, a later branching photoautotrophic clade in which the rTCA cycle was originally described~\cite{Buchanan90:rTCA}, ``Candidatus Nitrospira defluvii'', a member of the Nitrospirae in which the rTCA cycle was recently discovered through metagenome analysis~\cite{Lucker10:nitrospira}, as well as \emph{Sulfurimonas}, an $\epsilon$-proteobacterial family that uses the rTCA cycle~\cite{Hugler05:rTCA}, all use the less-symmetric rTCA variant. Their genomic features associated with the rTCA cycle are similar to those of the Hydrogenothermaceae and Desulfurobacteriaceae. All of these species possess a gene for ATP citrate lyase, have single copies of BC and CS enzymes, and have ICDH enzymes with a length of around 730--740 amino acid residues. 

Together these observations can be used to reconstruct a phylometabolic tree of rTCA variants, which in turn represents a branch of increased resolution on a more general phylometabolic tree of carbon-fixation~\cite{Braakman12:C_fixation}. This tree is shown in Fig.~\ref{fig:rTCA_tree}. The hypothesized root node, shown on the left, consists of the symmetric rTCA variant, but catalyzed by a single set of of enzymes for both arcs. Duplication and divergence toward greater specificity enzymes through selection for improved kinetics then gives rise to the symmetric rTCA variant found in Aquificaceae, while gene fusion in ATP citrate lyase combined with an increased substrate specificity of ICDH gives rise to a second divergence that leads to the conventional rTCA cycle. A comparison of ATP citrate lyase genes suggested that Aquificale families using it obtained this enzyme through HGT from $\epsilon$-proteobacteria, and that the Chlorobiale version represents the ancestral ATP citrate lyase~\cite{Hugler07:Aquifex_2TCAs}. Thus, the initial divergence between rTCA variants may have been between Aquificales and Chlorobiales, with two of the Aquificale families later joining the conventional rTCA group through HGT, as a result of ecological association with $\epsilon$-proteobacteria.

\begin{widetext}
\begin{center}
\begin{figure}[h]
\begin{center}
\includegraphics[scale=0.75]{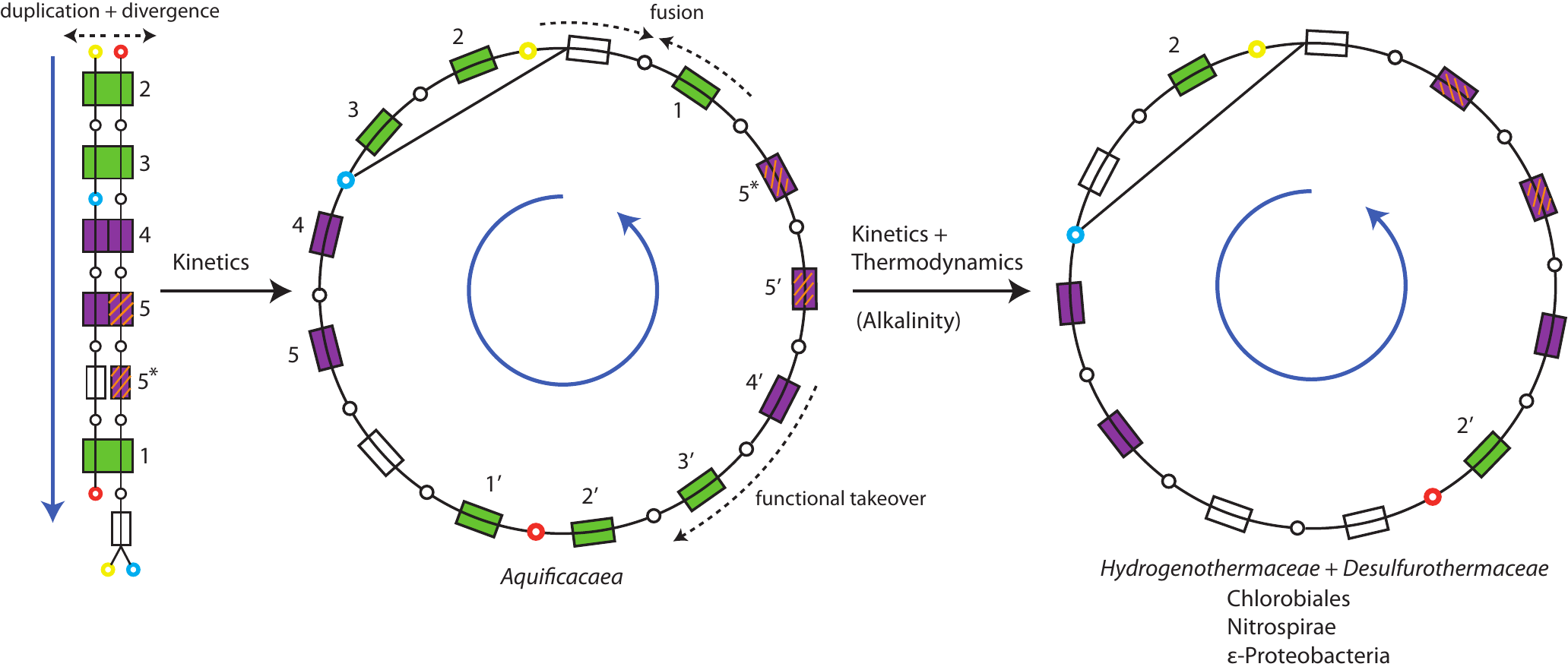}
\end{center}
\caption{Phylometabolic tree showing the evolution of the rTCA cycle. A combination of improving kinetics (which increases growth rate) and improving thermodynamics (which increases growth efficiency) explain both divergences. For the first divergence, duplication and divergence toward higher substrate specificity of enzymes improves kinetics. For the second divergence, replacing enzymes 3$^{\prime}$ + 4$^{\prime}$ by a higher specificity version of enzyme 4 removes a cost of 1 ATP hydrolysis in fixing CO$_2$, while fusion of one sub-unit of enzyme 1 with enzyme for the subsequent cleavage reaction improves kinetics. Green boxes represent homologous reactions catalyzed by enzymes with high sequence similarity, while purple boxes represent homologous reactions catalyzed by members of the same enzyme families. Reactions 5$^{\prime}$ and 5$^*$ are catalyzed by the same enzyme. Differences in sequence divergence between green and purple enzymes may reflect differences in complexity of the reactions, see text for further discussion. The yellow node represents acetyl-CoA, the blue node represents oxaloacetate, and the red node represents succinyl-CoA. Dark blue arrows indicate the direction of mass through pathways.}
\label{fig:rTCA_tree}
\end{figure}
\end{center}
\end{widetext}

While not all five homologous enzymes in the symmetric Aquificaceae rTCA variant show the same sequence similarity as the first three reaction (1--3), we argue this is consistent with the suggested evolutionary scenario. Reactions 1--3 represent thermodynamic bottleneck reaction that in isolation would require ATP hydrolysis~\cite{BarEven12:CFix_thermo}, and are catalyzed by elaborate multi-subunit enzymes that contain complex metal-centers and/or complex organic cofactors, and belong to small and highly conserved enzyme families. By contrast, the hydrogenation and dehydration reactions (4, 5) that follow these complex reactions are catalyzed by simpler enzymes belonging to highly diversified enzyme families used throughout metabolism. The selection pressure to conserve enzyme sequence similarity is therefore likely to have been lower, and either (or both) of these enzymes could also have been more easily replaced by other members of this family. This breakdown between ``easy'' and ``hard'' chemistries has been recognized as a general constraint in the early evolution of carbon-fixation pathways, with innovations primarily occurring through the emergence of new sequences of easier chemistries that connect harder bottleneck reactions~\cite{Braakman13:Logic}.

Having both arcs catalyzed by single sets of promiscuous enzymes in the root node would likely have resulted in slower growth for organisms using such a strategy, but it would also have considerably lowered the overhead for early regulatory machinery. If both rTCA arcs could have been catalyzed by the same set of single enzymes, the whole cycle would required only 7 total enzymes for catalysis: the 5 homologous reactions plus the reduction of malate to fumarate, and the cleavage of citryl-CoA, both of which do not have analogs in the opposite arc. It has previously been argued that a fully connected rTCA+WL network was selected as the root of carbon-fixation, because its topology would have provided the most robust form of network autocatalysis for earlier eras of life~\cite{Braakman12:C_fixation,Braakman13:Logic}. The present analysis suggests that an additional reason the rTCA cycle could have been privileged as a carbon fixation pathway capable of initiating the first cellular life, over alternate autocatalytic carbon-fixation pathways observed today~\cite{Hugler11:C_fix,Fuchs11:C_fix}, is that its greater symmetry permitted it to emerge with simpler catalytic support and regulatory structure.

Each of the divergences in the tree of rTCA variants can be understood in terms of a cost-benefit tradeoff in metabolic innovations. Duplication and divergence toward greater substrate specificity of the enzymes catalyzing the two arcs would have improved the kinetics of the pathway, and thus the growth rate of organisms using it. While the genome expansion due to these duplications would have had increased cost associated to it, this was apparently more than offset by the benefit of increasing the kinetics of a pathway through which most cellular carbon is incorporated. We will discuss a different case in the biosynthesis of branched chain amino acids in the next section, where a lower mass flux appears to have shifted this balance away from favoring duplication and divergence. 

For the second divergence, the fusion of the citryl-CoA synthase and citryl-CoA lyase would have improved the kinetics of the rTCA cycle, while the elimination of the ATP dependent $\alpha$-ketoglutarate BC would have improved the thermodynamic efficiency of the rTCA cycle. A secondary effect of this adaptation is that it replaces a carboxylation reaction based on HCO$_3^-$ with one based on CO$_2$, potentially allowing the new phenotype to thrive in less alkaline environments. It is interesting to note that a similar replacement of pyruvate BC and malate dehydrogenase, between whih oxaloacetate is the intermediate, in the opposite rTCA arc is not observed in any organism using rTCA to fix CO$_2$. This may simply be because oxaloacetate is the starting precursor to a wide range of anabolic pathways, and is most easily accessed if it is released into solution, and because it is moreover produced as a free intermediate during the cleavage of citrate.

The distribution of the different innovations governing the evolution of the two extant forms of the rTCA cycle may give some insights into the tradeoffs between optimizing growth rate and growth efficiency. The combined fitness advantage of improving kinetics by fusing the citrate cleavage sequence and improving thermodynamic efficiency through the emergence of the higher-specificity ICDH appears to be significant under most conditions. Only one family (Aquificaceae) within one clade (Aquificales) still uses the ancestral rTCA variant, while the conventional form is distributed across a wide range of bacterial clades. It is further interesting to note that in the vast majority of cases the two innovations occur together. In only one known case (\emph{S. azorense}) has the fusion of the citrate cleavage reaction taken place without elimination of the $\alpha$-ketoglutarate BC. Characterization of additional Aquificales could help us disentangle the ordering and relative advantage of the two innovations.

\subsection*{Amino acid biosynthesis}

For most amino acid biosynthetic pathways in \emph{A. aeolicus} the genome annotation leaves little doubt about the correct completion. Alanine, glutamate and aspartate are only one amination reaction removed from intermediates in the rTCA cycle, while an additional amination reaction leads to asparagine (from aspartate) and glutamine (from glutamate). The 3-step sequence from aspartate to homoserine provides the branching point from which the synthesis of threonine, methionine, and lysine (\emph{via} the DAP pathway~\cite{Scapin98:lysine}) diverge. Arginine and proline are both derive from glutamate, with arginine synthesis proceeding \emph{via} ornithine and a partial urea cycle. Histidine is synthesized from ATP and PRPP through the standard histidine synthesis pathway. The shikimate pathway~\cite{Bentley90:shikimate} produces chorismate, from which the syntheses of the aromatic amino acids tryptophan, phenylalanine and tyrosine diverge.

Surprising pathway variants used by \emph{A. aeolicus} for the synthesis of several amino acids were identified as a result of gaps in conventional pathways. Subsequent analysis further showed those pathways to represent the likely ancestral pathways to those compounds. In the previous section we mentioned the synthesis of glycine, serine, and cysteine, which in this organism are derived directly from CO$_2$ and NH$_3$ (and H$_2$S for cysteine) through a partial WL pathway coupled to the reductive (= biosynthetic) operation of the ``glycine cycle''~\cite{Braakman12:C_fixation}. Next we discuss the biosynthesis of the branched chain amino acids valine, leucine and isoleucine.

\subsubsection*{Branched chain amino acids}

It has long been thought that most organisms synthesize $\alpha$-ketobutyrate, a central precursor to isoleucine, by deaminating threonine~\cite{Umbarger57:Threonine}. However, an increasing number of species have been found to instead derive $\alpha$-ketobutyrate from pyruvate and acetyl-CoA through what is known as the ``citramalate'' pathway (see Fig.~\ref{fig:Branched_AAs}).  Originally discovered, and later described in detail, in the Spirochetes~\cite{Charon74:Ileu_Lepto,Xu04:Ileu_Lepto}, this pathway was subsequently discovered in a range of species across the bacterial and archaeal domains. It was found to be used in members of both the Euryarcheota~\cite{Eikmanns83:Ileu_Methano,Hochuli99:AA_Haloarcula,Howell99:Ileu_Methano,Drevland07:Ileu_Methano} and Crenarcheota~\cite{Schafer86:Carbon_Thermoprot,Jahn07:Carbon_Ignicoc}, as well as Firmicutes (Clostridia)~\cite{Feng09:Ileu_Firm,Tang10:Ileu_Helio}, Chloroflexi~\cite{Tang09:Ileu_Dehalo}, Chlorobia~\cite{Feng10:Ileu_Chlorobi}, Cyanobacteria~\cite{Wu10:Ileu_Cyano}, and several Proteobacteria~\cite{Risso08:Ileu_Geobac,Tang09:Ileu_Roseo,McKinlay10:Ileu_aProteo}.

Like the observation that the ancestral form of rTCA had an increased degree of symmetry that could have allowed a smaller regulatory structure, the sequence of local functional group transformations in the citramalate pathway is repeated in the synthesis of leucine from $\alpha$-ketoisovalerate (see Fig.~\ref{fig:Branched_AAs}). In contrast to the symmetric rTCA variant where parallel reactions are catalyzed by homologous enzymes, however, in this case the parallel reactions are in fact catalyzed by the same enzymes in both pathways~\cite{Xu04:Ileu_Lepto}. The only exceptions are the pair of acetyl-CoA addition reactions (5, 5$^{\prime}$) that initiate the two sequences, which are catalyzed by separate, though still homologous enzymes. Together with the broad observed distribution, this leads us to propose that the citramalate pathway represents the ancestral pathway to isoleucine, with threonine deaminase (often called threonine dehydratase, TDH) a more recent innovation.

The genome of \emph{A. aeolicus} is consistent with this hypothesis, as it lacks the gene for threonine deaminase and instead contains two genes annotated as 2-isopropyl malate synthase (IPMS, reaction 5$^{\prime\prime}$)~\cite{Deckert98:Aquifex_genome}. To confirm the presence of the citramalate pathway, and to place this use in evolutionary context, we performed broad genomic surveys for threonine deaminase (TDH) and citramalate synthase (CMS, reaction 5$^{\prime}$). Presence of either gene satisfies the constraint of pathway completeness in PMA, as the other genes necessary are shared between both pathways. Distributions of the two genes/pathways are shown in Table~\ref{tab:Ileu_genes}.

\begin{widetext}
\begin{center}
\begin{figure}[h]
\begin{center}
\includegraphics[scale=0.75]{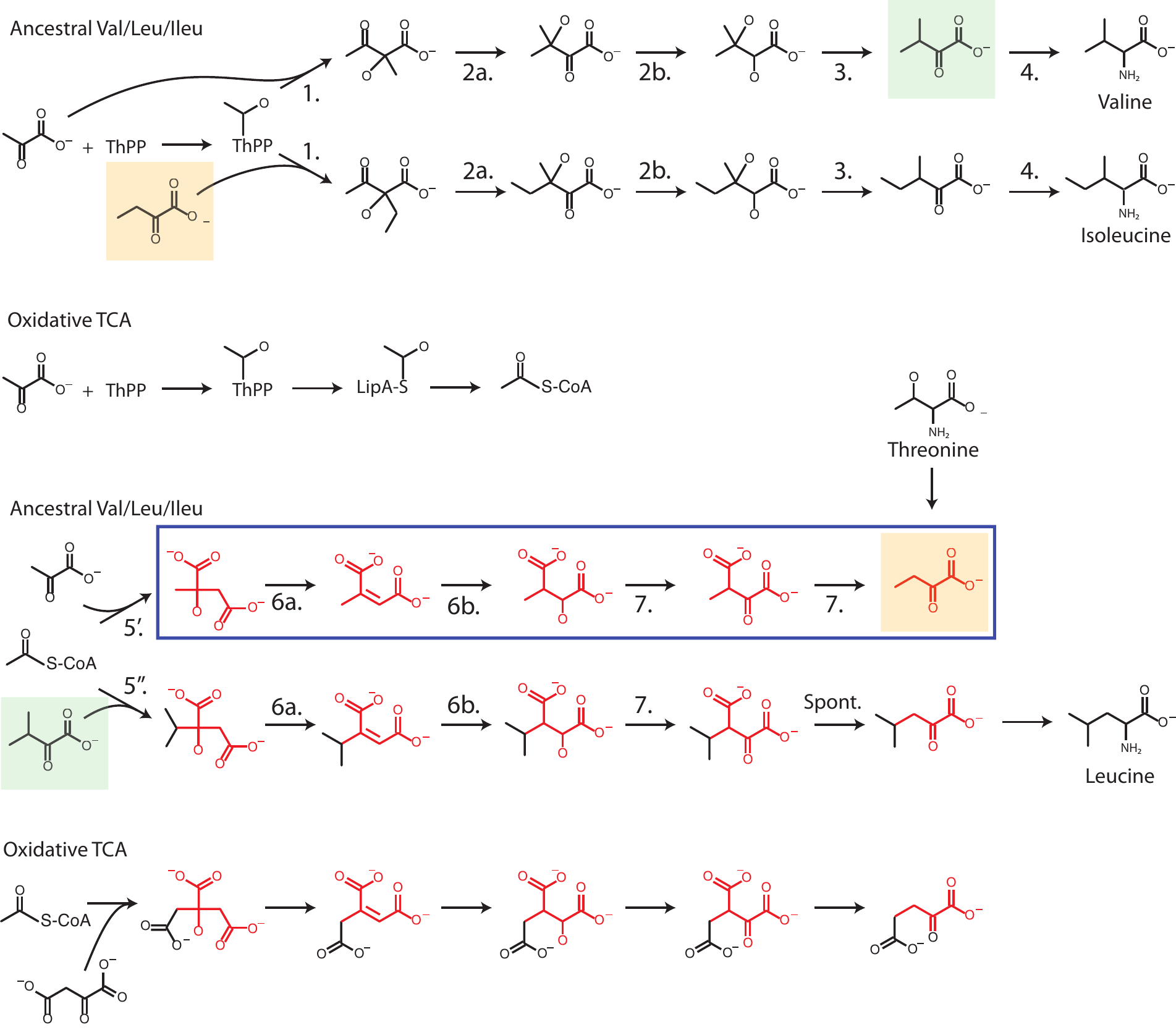}
\end{center}
\caption{Branched chain amino acid biosynthesis. The chemical sequences show the parallels in terms of local functional group chemistry within the reconstructed ancestral pathways to valine, leucine and isoleucine. The blue box highlight the citramalate  pathway of $\alpha$-ketobutyrate synthesis, reconstructed here to represent the ancestral sequence to this compound. The molecules highlighted in orange and green in turn show the compact interconnectedness of the ancestral pathways to the branched chain amino acids. Parallels to substrate sequences within the oxidative TCA are also highlighted, as well as the alternate route to $\alpha$-ketobutyrate from threonine.}
\label{fig:Branched_AAs}
\end{figure}
\end{center}
\end{widetext}

We note briefly that there is some uncertainty in the annotation of CMS, because of the sequence similarity not only to IPMS, but also to homocitrate synthase of the AAA Lysine synthesis pathways~\cite{Drevland07:Ileu_Methano}. However, the pooling of evidence in PMA lowers these uncertainties. For example, if a strain has two genes annotated as IPMS, possesses no gene for TDH, nor any of the other genes of the AAA lysine pathway to which homocitrate synthase might connect, one of the IPMS copies most likely instead codes for CMS. In addition, BLAST search comparisons among these genes were used for cross-validation. For most species with multiple genes annotated as IPMS, BLAST searches show that at the clade level these genes separate into clear groups based on sequence similarity. For most clades, genes within one of these groups were identified in the various experimental studies mentioned above as likely encoding CMS, thus anchoring the group as a whole. Combining all these different lines of evidence leads to the distributions shown in Table~\ref{tab:Ileu_genes}.

\begin{table}[!ht]
\footnotesize
\begin{tabular}{|l|lll|}
\hline
  &  TDH & CMS & both \\
  Domain/Clade/Family & & &\\
\hline
 Archaea (122)  & 58 & 98 & 47\\ 
 ~~Crenarcheota (40)  & 34 & 30 & 27\\
 ~~Euryarcheota (79)  & 21 & 65 & 17\\
 ~~Korarcheota (1)  & 1 & 1 & 1\\
 ~~Thaumarcheota (2)  & 2 & 2 & 2\\
\hline
 Bacteria total (321) & 213 & 138 & 55\\
 Bacteria w/o Firmicutes (118) & 57 & 78 & 29\\
 ~~Aquificales (9) & 0 & 9 & 0\\
 ~~Thermotogales (13)  & 5 & 6 & 5\\
 ~~Chlorobiales (11) & 0 & 11 & 0\\
 ~~Chloroflexi (16) & 10 & 12 & 6\\
 ~~Nitrospirae (3) & 0 & 3 & 0\\
 ~~Planctomycetes (6) & 1 & 6 & 1\\
 ~~Verrucomicrobia (4) & 1 & 4 & 1\\ 
 ~~Deinococcus-Thermus (15)  & 10 & 9 & 9\\
 ~~~\emph{Deinococcales} (7) & 6 & 1 & 0\\
 ~~~\emph{Thermales} (8) & 4 & 8 & 4\\
 ~~Cyanobacteria (41) & 30 & 21 & 12\\
 ~~~\emph{Prochlorococcus} (12) & 12 & 0 & 0 \\
 ~~~\emph{Synechococcus} (11) & 8 & 5 & 2 \\
 ~~~Others (18) & 10 & 16 & 10 \\
 ~~Firmicutes (203) & 156 & 60 & 26 \\
 ~~~Bacilli (104)  & 101 & 9 & 9\\
 ~~~Clostridiales (99) & 55 & 51 & 17\\
 ~~~~\emph{Clostridiaceae} (33) & 26 & 8 & 3\\
 ~~~~\emph{Thermoanaerobacter} (24) & 3 & 20 & 1\\
\hline
\end{tabular}
\caption{
  Distribution of Isoleucine biosynthesis pathways.
\label{tab:Ileu_genes}
}
\end{table}

Among the archaea in our study, a vast majority (98/122, or 80\%) posses a gene for CMS. By contrast, less than half (58/122, or 47.5\%) possess a gene for TDH. Moreover, when TDH is found it is mostly co-present with CMS, while in many cases CMS represents the exclusive pathway to isoleucine. This is relevant, because two versions of TDH are known to exist: an `anabolic' version whose activity is obligatory for cell growth, and a `catabolic' version that only becomes active during salvage of excess threonine~\cite{Umbarger57:Threonine}. It has previously been noted that the archaeal TDH best matches the catabolic version of this enzyme~\cite{Xie02:Tryptophan}. This is consistent with experimental observation that for archaea which have both TDH and CMS, the citramalate pathway represent either a major or the exclusive pathway used in synthesis of isoleucine~\cite{Schafer86:Carbon_Thermoprot,Hochuli99:AA_Haloarcula}. Thus, we conclude that the citramalate pathway represents the ancestral pathway to isoleucine in archaea, with TDH representing a later innovation that initially emerged for salvage purposes.

Bacteria present a more complex picture than archaea regarding the synthesis of isoleucine. Both TDH and CMS are common in deep-branching bacterial clades, with TDH occurring with higher frequency.  However, this balance is dominated by Firmicutes, and among other deep-branching bacteria, the citramalate pathway appears to be the most abundant route to isoleucine. Strikingly, a significant number of deep-branching clades whose members include (hyper)thermophiles, autotrophs, or both, show near exclusive use of the citramalate pathway: Aquificales, Chlorobiales, Nitrospirae, Planctomycetes, and Verrucomicrobia.

A closer look within clades where TDH represents the majority pathway provides further context for evolutionary reconstruction. For example, within the large and diverse Firmicute phylum, the aerobic Bacillales nearly all possess the threonine pathway, while the anaerobic Clostridiales exhibit the threonine and citramalate pathways in about equal frequencies. Within the Clostridiales in turn, the Clostridiaceae family, which contains many pathogenic strains~\cite{Wiegel06:clostridium}, make nearly exclusive use of the threonine pathway, while the Thermoanaerobacteriales family, which like the Aquificales are hyperthermophiles, make nearly exclusive use of the citramalate pathway. 

Within the Cyanobacteria in turn, the majority of strains using the threonine pathway belong to the genera \emph{Synechococcus} and \emph{Prochlorococcus}. Many of the former and all of the latter are divergent cyanobacterial lineages highly adapted to oligotrophic regions of the world's oceans, often dominating those environments~\cite{Urbach98:ProchSynech_diverge,Palenik03:Synechococcus,Rocap03:Prochlorococcus}. Among all other cyanobacteria, the citramalate pathway represents the major route to isoleucine. Finally, within Deinococcales-Thermales~\cite{Omelchenko05:deinococcales}, which have approximately equal numbers of both pathways, the Deinococcales use mainly the threonine pathway, while the hyperthermophilic Thermales use mainly the citramalate pathway.

Lastly, it has previously been observed that many deep-branching bacteria that possess both TDH and CMS, appear to possess the catabolic version of TDH~\cite{Xie02:Tryptophan}. In our sample of Thermotogales and Chloroflexi, both of which show frequent occurrence of TDH, BLAST searches show that many of these genes are better matched to the catabolic than the anabolic TDH of \emph{E.  coli}. Thus, all evidence combined leads us to conclude that the citramalate pathway is the ancestral pathway also in bacteria, and thus for all life, with the threonine pathway initially emerging as a salvage pathway.

This conclusion is important relative to the previous discussion of a minimal ancestral rTCA cycle. In addition to the enzymes shared between the citramalate pathway and the final sequence in leucine synthesis (reactions 5--7), the enzymes catalyzing the homologous sequences in valine and isoleucine synthesis (reactions 1--3) are similarly shared across pathways, while the final amination reaction (4) is in all three pathways performed by the same enzyme~\cite{Umbarger78:AA_synthesis}. While some organisms have additional copies of the first thiamin catalyzed reaction (1) for purposes of regulation~\cite{Grimminger79:branchedAAs,Barak87:branchedAAs}, \emph{A. aeolicus} possesses only a single, two-subunit enzyme for both reactions~\cite{Deckert98:Aquifex_genome}. As mentioned, the lone exception to this pattern of promiscuous catalysis is the acetyl-CoA addition reaction (5, 5$^{\prime}$) initiating both sequences in leucine/isoleucine synthesis. However, if these reaction had not been mediated by thioester intermediates, they would represent carboxyl reduction and carbon-carbon bond forming reactions that in carbon-fixation pathways are associated with ATP hydrolysis~\cite{BarEven12:CFix_thermo}. Similar to those more constrained carbon-fixation reactions, the high sequence similarity of the two enzymes caused them to be originally annotated as the same enzyme~\cite{Deckert98:Aquifex_genome,Xu04:Ileu_Lepto}. The more complex, constrained nature of these reactions may thus have increased the kinetic advantage of duplication and divergence toward greater enzyme specificity. Thus, if as before we assume an ancestral enzyme with broader substrate affinity for reactions 5/5$^{\prime}$, then all 21 total reactions in the synthesis of valine, leucine and isoleucine starting from pyruvate and acetyl-CoA would have required only 7 enzymes for catalysis (see also Fig.~\ref{fig:Branched_AAs}).

In contrast to the evolution of the rTCA cycle, most enzymes catalyzing homologous reaction sequences in this case have not duplicated and diverged, except for CMS and IMPS. Why this difference? The difference in mass flux between the two pathways would appear to a offer a straightforward explanation. While most cellular carbon passes through the rTCA cycle in autotrophs using that pathway, only 3 out of 20 amino acids are generated through the pathways in Fig.~\ref{fig:Branched_AAs}. Thus, even if we simplistically assume roughly equal biomass partitioning between amino acids, nucleotides and lipids, the mass flux difference between the two sets of pathways is about an order of magnitude. The only enzyme for which selection pressure to improve kinetics appears to have led to duplication and divergence within the synthesis of branched chain amino acids represents a more complex, possible thermodynamic bottleneck reaction.

\subsubsection*{Emergence, and combined regulation with the rTCA cycle}

Additional overlaps between the rTCA cycle and the branched chain amino acid biosynthetic pathways may provide insights into how the latter emerged. The dehydration/hydration isomerization and dehydrogenation sequences (reactions 6a, 6b, 7) in the reconstructed ancestral pathways to leucine and isoleucine is homologous to the sequence of local group transformations occurring in the opposite direction in the large-molecule half of rTCA (reactions 4$^{\prime}$, 5$^{\prime}$ and 5$^*$ in Fig.~\ref{fig:C_Fix}). The associated enzymes in the two sub-systems may also have a common origin. In \emph{A. aeolicus} (which uses the ancestral rTCA variant), the dehydration/hydration-isomerization reaction is catalyzed by a single subunit aconitase enzyme (ACO) in rTCA, and a two subunit enzyme (LeuC/D) in the branched chain amino acid pathways. However, the combined length of LeuC and LeuD is similar to that of ACO, while LeuC and LeuD also have high sequence similarity to consecutive, adjacent portions within ACO (data not shown). This suggest the possibility that following duplication and divergence from a common ancestral enzyme the two subunits became fused within the rTCA cycle but not the branched chain amino acid pathways, due to the differences in mass flux density. Similarly, the homologous (de)hydrogenation reactions in the two sub-systems are in \emph{A. aeolicus} also catalyzed by enzymes with high sequence similarity. The enzyme homologies do not extend to the (de)carboxylation reactions across the two sub-systems, which in the direction of decarboxylation is facile, and in the direction of carboxylation requires complex cofactors and ATP hydrolysis (see previous discussions). Phylogenetic reconstructions of the lineages of these enzymes could shed additional light on these hypotheses.

These observation may thus suggest that the existence of a sequence of substrate chemistry operating in the opposite direction within the ancestral rTCA cycle could have facilitated the emergence of the ancestral pathways to leucine and isoleucine. The only truly new form of chemistry in the citramalate pathway and its homologous sequence in leucine synthesis is the initiating reaction involving ligation of acetyl-CoA (reactions 5, 5$^{\prime}$ in Fig.~\ref{fig:Branched_AAs}). Moreover, the facile decarboxylation of $\beta$-carboxylic acids, which in the reductive direction of the rTCA cycle may have increased the cost of an already complex reaction, may have created an advantage when used in the opposite direction, possibly further facilitating the emergence of the ancestral citramalate/leucine sequence.


The pattern of pathway diversification by innovation of the initiating reaction, followed by re-use of similar or identical downstream enzymes to catalyze homologous reaction sequences, was also the primary mode of diversification proposed for carbon-fixation pathways in Ref.~\cite{Braakman13:Logic}. The difference to those previous proposals is that here also the direction of the re-used sequence chemistry is suggested to have changed. We should also note that in comparing these innovations some caution should be exercised because the establishment of the first pathway to a set of amino acids is an innovation occurring in the era prior to LUCA, while diversification of carbon-fixation pathways occurred after LUCA. Nonetheless the parallels in the suggested modes of innovations is striking and may suggest a general principle of early metabolic evolution.

If true, this hypothesis of pathway evolution may have allowed the LUCA to regulate the combined sub-systems with an even smaller genome than we have suggested for them separately. If the homologous enzymes in the two sub-systems arose through duplication and divergence from a common ancestor, then the two sequences could have potentially been catalyzed by the same enzymes in an earlier era. We suggested above that the complete ancestral forms of rTCA and branched chain amino acid biosynthesis could in an era of more promiscuous enzymes each have been catalyzed by only 7 enzymes total. The added observations here suggest that the entire connected network of rTCA plus the pathways to the branched chain amino acids could have been catalyzed by only 12 total enzymes.

\subsubsection*{Relation to the reversal of the TCA cycle}

The key innovation of the acetyl-CoA ligation reaction that initiates the homologous sequences within leucine and isoleucine synthesis, and possibly governed their emergence, may have also facilitated the later emergence of other pathways within metabolism, in particular the reversal of direction of the TCA cycle. The entire homologous reaction sequences (5--7) in the citramalate and leucine pathways are also known as ``keto acid elongation'' sequences~\cite{Jensen76:recruitment}, and are further repeated within the oxidative TCA cycle (as well as the AAA lysine synthesis pathway). It is becoming clear that the TCA cycle originally operated in the reductive direction (see Ref.~\cite{Braakman13:Logic} and references therein for discussion), which means that keto acid elongation likely occurred within branched chain amino acid synthesis prior to its use in the oxidative TCA cycle. The thiamin-facilitated decarboxylation of pyruvate (see Fig.~\ref{fig:Branched_AAs}) thus similarly appears to have been used first in the synthesis of branched chain amino acids then in its use within the oxidative TCA cycle. Finally, lipoic acid is used only in the oxidative (and not the reductive) direction of the TCA cycle, while its likely ancestral function was in the synthesis of glycine through the glycine cycle~\cite{Braakman12:C_fixation} (see also the section on lipoic acid below for additional supporting evidence).

Thus, we suggest that at the substrate level the prior existence of the biosynthetic pathways leading to valine/leucine/isoleucine facilitated the reversal of the TCA cycle from the reductive to the
oxidative direction. The additional key innovation appears to have been the recruitment of lipoic acid from the glycine cycle to its interaction with thiamin in the production of acetyl-CoA from pyruvate. Broad affinity of earlier enzymes would have aided the emergence of this novel pathway as promiscuous activity could have allowed this pathway to proceed at lower rates, with duplication and divergence later being favored as respiration came fully online and mass flux through this pathway increased.

\subsection*{Cofactors}

Cofactors are a distinct class of molecules at the substrate level of metabolism, forming a chimeric intermediary layer between monomers and polymers in terms of structure~\cite{Srinivasan09:aquifex_chart}. Cofactors are also critical components of the control hierarchy of metabolism, facilitating many key reaction mechanisms, and thus the overall integration of metabolism. Each cofactor generally facilitates a distinct and specialized catalytic function, and their emergence can thus be thought of as the outgrowth of kinetic feedback loops, each ``opening up'' new spaces in the universe of organic chemistry and bringing them under the control of biology~\cite{Braakman13:Logic}. Understanding the evolution of cofactor biosynthesis is thus important both in providing context to discussions on the origin of life, as well as understanding major physiological lineages in the tree of life. In this section we focus on the synthesis of several cofactors in \emph{A. aeolicus}, using the reconstruction to provide additional insights into the evolution of their functions.

\subsubsection*{Lipoic acid}

Lipoic acid is a cofactor with very limited, but key metabolic roles. Lipoic acid is central to the ``Glycine Cleavage System'' (GCS)~\cite{Kikuchi73:GCS}, which connects glycine and serine metabolism to folate one-carbon chemistry, and is also used (as previously mentioned) in the ferredoxin:oxidoreductase decarboxylation reactions in the oxidative TCA cycle and the degradation of branched chain amino acids. The GCS is known to be reversible~\cite{Kikuchi73:GCS}, has nearly neutral thermodynamics~\cite{BarEven12:CFix_thermo}, and likely originally operated in the reductive (i.e. biosynthetic, not degradative) direction as part of the ancestral pathway leading to glycine and serine~\cite{Braakman12:C_fixation}. For these reasons the GCS together with serine methyl transferase (SMT) has also been called the ``Glycine Cycle''~\cite{Braakman12:C_fixation}.  The phylometabolic analysis that places reductive glycine synthesis at the base of the tree of life, as part of the phenotype of the last common ancestor, suggests that the function of lipoic acid in the glycine cycle preceded its use in either the oxidative TCA cycle or the degradation of branched chain amino acids. This provides important context for interpreting the distribution of lipoic acid biosynthesis genes that we discuss next.

\begin{figure}[h!]
\begin{center}
\includegraphics[scale=1]{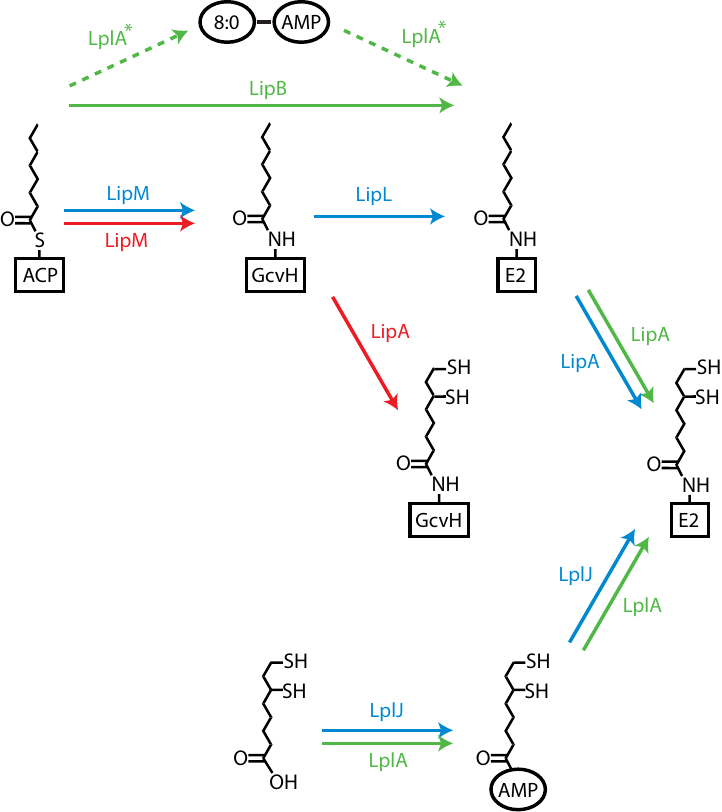}
\end{center}
\caption{Lipoic acid biosynthesis and lipoyl-protein assembly. In \emph{E. coli} (green sequence), octanoate is transfered from ACP to the E2 subunit of pyruvate dehydrogenase (PDH) by LipB, followed by sulfuration to lipoic acid by LipA. In \emph{E. coli} mutants lacking LipB, octanoate is transfered through an alternate route with an AMP-bound intermediate by LplA, normally used for incorporation of free lipoic acid. In \emph{B. subtillis} (blue sequence), octanoate is transfered from ACP first to the H-protein of GCS by LipM, followed by a second transfer to the E2 subunit of PDH by LipL. \emph{B. subtillis} also uses LplJ instead of LplA for incorporation of free lipoic acid. In red is the suggested ancestral biosynthesis of lipoic acid (see main text).}
\label{fig:Lipoate}
\end{figure}

Three pathways are known for the biosynthesis of lipoic acid (see Fig.~\ref{fig:Lipoate}). The conventional pathway, first described in \emph{E. coli}~\cite{Cronan05:LA_Rev}, involves transfer of an octanoyl moiety from the Acyl Carrier Protein (ACP) to one of the Lipoyl Dependent (LD) enzymes, followed by sequential sulfuration of the octanoyl moiety to produce the final lipoated enzyme~\cite{Zhao03:LA_attach}. The first step is catalyzed by octanoyl transferase (LipB), while the second is catalyzed by Lipoyl Synthase (LipA). A variant of this scheme was recently discovered in \emph{B. subtillis}, which involves the same basic chemistry, but distinct set of enzymes and an additional intermediate in the transfer of octanoate. In \emph{B. subtillis}, the distinct octanoyl transferase LipM transfers octanoate from ACP to the H-protein of the GCS, followed by a second transfer (catalyzed by LipL) to the E2 subunit of pyruvate dehydrogenase~\cite{Christensen10:Octanoyl,Martin11:LipoicAcid,Christensen11:LipoicAcid}. Both LipM and LipL had previously been obscured due to their sequence similarities to LplA of \emph{E. coli} ~\cite{Martin11:LipoicAcid}. A third distinct pathway was recently discovered in an \emph{E. coli} mutant in which LipB had been deactivated. In this mutant, lipoate protein ligase (LplA), normally used in the attachment of free lipoic acid to LD enzymes, is recruited in the transfer of free octanoate to an LD enzyme through an AMP-bound intermediate~\cite{Zhao03:LA_attach,Booker04:LA_new,Hermes09:LipoicMutant,Rock09:LipoicNew}. In light of these variations, it is noteworthy that \emph{A. aeolicus} lacks a gene for LipB and is annotated as having LipA and LplA~\cite{Deckert98:Aquifex_genome}. This raises the question, which of the pathway alternates does \emph{A. aeolicus} use, and what does this teach us about how the biosynthetic pathways and the functionality of lipoic acid evolved?

To examine this question, we performed broad genomic surveys for each of the enzymes used in lipoic acid metabolism, shown in Table~\ref{table:lipoate}. In addition to LipA, LipB, LplA, LipM/LipL, our survey also includes LplJ, a distinct lipoate protein ligase found in \emph{B. subtillis}~\cite{Martin11:LipoicAcid}. It can immediately be seen that the conventional LipA + LipB combination is not widely distributed across deep-branching clades. Only 6 archaeal strains, nearly all (5/6) in the Thermoproteale family within the Crenarcheota, appear to possess this pathway. Among deep-branching bacteria, only the Deinococcales, Cyanobacteria and Clostridiale family within Firmicutes show significant abundance of the LipA + LipB combination. In contrast, LplA (and its combination with LipA) is widely and more evenly distributed across both archaea and bacteria. However, BLAST searches indicate that in most of these cases LplA is in fact a better match to LipM of \emph{B. subtillis} than to LplA of either \emph{E. coli} or the euryarchaeota \emph{T. acidophilum}~\cite{Christensen09:LA_Thermoplasma}. While this suggests that the \emph{B. subtillis} variant may be the ancestral pathway to lipoic acid, in most cases with a putative LipM gene, we could not find the accompanying LipL gene, presenting a puzzle as this gene is absolutely required in lipoic acid synthesis in \emph{B. subtillis}~\cite{Martin11:LipoicAcid}.

This is where the functional roles of lipoic acid provides critical evolutionary context. As explained, the likely ancestral function of lipoic acid is its role in connecting glycine/serine metabolism to folate-C$_1$ chemistry through the glycine cycle, for which it remains (nearly universally) essential to this day. In contrast, the role of lipoic acid in the oxidative TCA cycle or the degradation of branched chain amino acids likely arose later, and is not essential to many organisms. For example, many autotrophs, including \emph{A. aeolicus}, do not use the oxidative TCA cycle nor do they degrade branched chain amino acids. Such organisms should thus not need LipL to transfer octanoate from the H-protein of GCS to other LD enzymes, because they do not possess them. Instead, direct sulfuration of octanoate bound to the H-protein of GCS, a reaction known to be catalyzed by LipA~\cite{Cicchillo04:LipA}, is sufficient to allow the sole function of lipoic acid in these organisms. This proposed sequence of enzyme functions is supported by the observation that in deep-branching clades that do not possess the glycine cycle -- including many Clostridia, all methanogenic families within the Euryarcheota or the Desulfurobacteriaceae family within Aquificales~\cite{Braakman12:C_fixation} -- we do not find any genes associated with lipoic acid biosynthesis or uptake. Thus, we suggest that in many cases where a putative LipM is found but LipL is absent, a pathway is used in which LipM is directly followed by LipA (see Fig.~\ref{fig:Lipoate}), and we also propose that this represents the ancestral pathway for \emph{de novo} lipoic acid biosynthesis. Note that this proposal further reinforces the conclusion that the reductive TCA cycle preceded the oxidative version.

\begin{widetext}
\begin{center}
\begin{table}[ht]
\footnotesize
\begin{tabular}{|l|lllllllll|}
\hline
  & LipA & LipB & A+B & LplA & A+A & LipM & LipL & A+L+M & LplJ\\
  Domain/Clade/Family & & & & & & & & & \\
\hline
  Archaea total (126) & 37 & 6 & 6 & 60 & 31 & & & &\\
  ~~Crenarcheota (41) & 18 & 5 & 5 & 27$^*$ & 13 & & & & \\
  ~~Euryarcheota (80) & 19 & 1 & 1 & 32 & 18 & & & &  \\
  ~~Korarcheota (1) & 0 & 0 & 0 & 1$^*$ & 0 & & & & \\
  ~~Nanoarchaea (1) & 0 & 0 & 0 & 0 & 0 & & & & \\
  ~~Thaumarcheota (3) & 0 & 0 & 0 & 0 & 0 & & & & \\
\hline
  Bacteria total (329) & 214 & 105 & 102 & 100 & 83 & 94 & 108 & 85 & 146\\
  ~~Aquificales total (9) & 7 & 3 & 3 & 4$^*$ & 4 & & & & \\
  ~~~\emph{Aquificaceae} (4) & 4 & 0 & 0 & 4$^*$ & 4 & & & & \\
  ~~~\emph{Hydrogenobacter} (3) & 3 & 3 & 3 & 0 & 0 & & & & \\
  ~~~\emph{Desulfotobacter} (2) & 0 & 0 & 0 & 0 & 0 & & & & \\
  ~~Thermotogales (13) & 0 & 0 & 0 & 13$^*$ & 0 & & & & \\
  ~~Deinococcus-Thermales (15) & 15 & 15 & 15 & 2 & 2 & & & & \\
  ~~Chloroflexi (16) & 8 & 6 & 6 & 10$^*$ & 8 & & & & \\
  ~~Chlorobi (11) & 11 & 1 & 1 & 11$^*$ & 11 & & & & \\
  ~~Planctomycetes (6) & 5 & 1 & 0 & 5$^*$ & 5 & & & & \\
  ~~Nitrospira (3) & 2 & 0 & 0 & 3$^*$ & 2 & & & & \\
  ~~Verrucomicrobia (4) & 4 & 3 & 3 & 4$^*$ & 4 & & & & \\
  ~~Cyanobacteria (44) & 43 & 43 & 43 & 43$^*$ & 43 & & & & \\
  ~~Firmicutes total (208) & 119 & 33 & 31 & 5 & 4 & 94 & 108 & 85 & 146\\
  ~~~\emph{Bacilli} (106) & 92 & 7 & 7 & 0 & 0 & 85 & 106 & 85 & 96\\
  ~~~\emph{Clostridia} (102) & 27 & 26 & 24 & 5 & 4 & 9 & 2 & 0 & 50\\
  \hline
\end{tabular}
\caption{
Distribution of lipoic acid biosynthesis genes. Within the column counting LplA genes, entries denoted by ($^*$) are often better matched to LipM of \emph{B. subtillis} than LplA of either \emph{E. coli} or \emph{T. acidophilum}. See main text for additional details.
\label{table:lipoate}
}
\end{table}
\end{center}
\end{widetext}

In this scenario, an additional octanoyl transferase (LipL) emerged for the second subsequent transfer to the E2 domain of PDH, possibly through duplication and divergence of LipM as seen in \emph{B. subtillis}. This would have introduced a redundancy into the lipoic acid system by producing two dedicated octanoyl transferase enzymes. This redundancy then appears to have been removed by replacing the two distinct transferase enzymes with a single all-purpose LipB transferase, as seen in \emph{E. coli}. The high sequence similarity among LipM, LipL, LplA and LpIJ further suggests that the environmental uptake genes likewise arose through duplication and divergence from octanoyl transferase genes, but that some of this ancestral function was maintained in LplA, making possible its recruitment in \emph{E. coli} mutants lacking LipB.

\subsubsection*{Vitamin B6}

In contrast to the narrow functionality of lipoic acid, vitamin B6, which refers to pyroxidal 5-phosphate and its substitutes, is one of the most functionally diverse cofactors, with its different forms facilitating a very wide range of reaction classes~\cite{Fitzpatrick07:B6_review}. Its diverse functionality and relatively simple chemistry, plausibly accessible under abiotic conditions, has led to the suggestion that it may have been one of the earliest cofactors~\cite{Austin99:B6_prebiotic,Morowitz06:B6}. 

There are two known biosynthetic pathways leading to vitamin B6 in modern metabolism. In the first recognized pathway, described in \emph{E. coli}, pyridoxine phosphate is derived from 4-erythrophosphate~\cite{Fitzpatrick07:B6_review}. This pathway is known as the `DXP dependent' pathway, because 1-deoxy-D-xylulose-5-phosphate (DXP) is the secondary input to the final condensation reaction that produces the pyridine ring of pyridoxine. In the second (`DXP independent') pathway, first described in \emph{B. subtillis}, pyridoxal phosphate is synthesized through the direct condensation of ribulose phosphate and glyceraldehyde phosphate~\cite{Burns05:B6_2nd,Raschle05:B6_2nd}.

Previous analyses found genes for the DXP-independent pathway to be highly conserved and distributed across both archaeal and bacterial domains, while genes for the DXP-dependent pathway were found mainly in the $\gamma$-proteobacteria, suggesting this latter pathway emerged later in evolution~\cite{Ehrenshaft99:B6_evol,Mittenhuber01:B6_evol}. However, \emph{A. aeolicus} possesses the DXP-dependent pathway, prompting us to further examine this hypothesis. Table~\ref{tab:B6_genes} shows the distribution of the key enzymes involved in the condensation steps in both sequences -- PdxA/J in the DXP-dependent pathway and PdxS/T in the DXP-independent pathway -- across bacteria and archaea.

\begin{table}[ht]
\footnotesize
\begin{tabular}{|l|lll|lll|}
\hline
& \multicolumn{6}{|l|}{Pdx}  \\
\cline{2-7}
Domain/Clade/Family & S & T & S+T & A & J & A+J \\
\cline{1-7}
Archaea (126) & 114 & 114 & 113 & 6 & 0 & 0 \\
\hline
Bacteria (339) & & & & & & \\
~Aquificales (9) & 0 & 0 & 0 & 8 & 9 & 8 \\
~Thermotogales (13) & 12 & 12 & 12 & 0 & 0 & 0 \\
~Chloroflexi (16) & 16 & 15 & 15 & 0 & 0 & 0 \\
~Planctomycetes (6) & 0 & 0 & 0 & 6 & 6 & 6 \\
~Nitrospirae (3) & 0 & 0 & 0 & 3 & 3 & 3 \\
~Chlorobia (11) & 0 & 0 & 0 & 11 & 11 & 11 \\
~Deinococci (15) & 15 & 15 & 15 & 0 & 0 & 0 \\
~Verrucomicrobia (4) & 0 & 0 & 0 & 4 & 4 & 4 \\
~Cyanobacteria (44) & 0 & 0 & 0 & 44 & 44 & 44 \\
~Firmicutes (208) & 183 & 164 & 164 & 47 & 0 & 0 \\
~~~Bacillales (106) & 106 & 106 & 106 & 11 & 0 & 0 \\
~~~Clostridia (102) & 77 & 58 & 58 & 36 & 0 & 0 \\
\hline
\end{tabular}
\caption{
Distribution of pyroxidal phosphate synthesis genes. 
\label{tab:B6_genes}
}
\end{table}

These distributions show some striking patterns. It had previously been noted that the two pathways are mutually exclusive within organisms, which use only one or the other~\cite{Ehrenshaft99:B6_evol}. Our analysis shows that the pathways are in fact mutually exclusive at the \emph{clade} level, much more so than we have seen for pathway variants in other sub-systems we have previously analyzed. PdxA is found in a few species within both archaea and Firmicutes, but that enzyme catalyzes a hydrogenation reaction, with the enzyme catalyzing the actual ring condensation reaction (PdxJ) completely absent in those cases. Our analysis further appears to confirm that the DXP-independent pathway represents the ancestral pathway. In addition to nearly all archaea, several deep-branching bacterial clades (Thermotogales, Chloroflexi, Deinococcales, Firmicutes) also use this pathway.

It was previously suggested that the DXP-dependent pathway arose within proteobacteria~\cite{Mittenhuber01:B6_evol}, but the fact that all members of several deep-branching clades use this pathway suggests it may actually have been an earlier innovation. The observed distribution of both pathways is difficult to explain, however. That the pathways are mutually exclusive at the clade level requires either early HGT to progenitors of clades, extensive gene transfer between select clades after they had diverged, or extensive transfer within clades that can take the appearance of genes sweeping through the population~\cite{Polz13:HGT_eco}. In any of these cases there is no obvious explanation for why transfer would have been restricted to occurring only between select clades, nor is a selective advantage
apparent.

An alternative explanation would be (possibly repeated) convergent evolution early in the divergences of clades. This explanation has some appeal, as the key enzymes in both pathways, PdxA/J and PdxS/T, are in fact very similar in their 3-D structure and in the sequence of local functional group transformations that make up the respective condensation reactions~\cite{Fitzpatrick07:B6_review}. However, even for convergent evolution we are lacking a good explanation for why only some clades would pervasively adopt this new strategy, while others did not.

If the evolutionary sequences and driving forces are not clear, we can at least identify features of the two pathways that would have facilitated the transition between them. The shared fold structure and similarity in reaction mechanisms, but low sequence similarity, between PdxA/J and PdxS/T have been interpreted to mean that they represent convergent discoveries~\cite{Fitzpatrick07:B6_review}. However, it is also possible that PdxA/J emerged from PdxS/T and that both have been under strong selection pressure, causing their sequence to diverge strongly. 

Another common feature between pathways is they both start from intermediates within the pentose phosphate pathway. The key exception is DXP itself, which is the product of the first committed reaction in the DOXP pathway of terpenoid backbone synthesis. Archaea exclusively use the alternate mevalonate (MVA) pathway to synthesize terpenoids, while bacteria use both the MVA and DOXP pathways~\cite{Boucher00:terpenoids}, possibly providing a partial explanation for why the DXP-dependent pathway to vitamin B6 emerged only within bacteria and not archaea.

Finally, the other reactions in the DXP-dependent pathway that lead up to the condensation sequence catalyzed by PdxA/J represent common and widely used metabolic reactions catalyzed by members of highly diversified enzyme families. The reaction sequence connecting 4-erythrophosphate to phospho-4-hydroxy-threonine, the input to the PdxA/J-catalyzed ring condensation sequence, consists of a hydration/reduction of an aldehyde to a carboxyl group, a subsequent dehydrogenation of an alcohol to a carbonyl group, and finally the reductive amination of that carbonyl group. Especially in an earlier era of more promiscuous catalysts, this pathway could thus well have been recruited \emph{en bloc} into the emergent DXP-dependent pathway. The selective advantage of this adaptation, and the way it might have led to the peculiar distribution of both pathways, remains to be explained, however.

\subsubsection*{Quinones}

The main component of the quinone pool in \emph{A. aeolicus} was determined to be 2-demethylmenaquinone-7 (DMK-7)~\cite{Infossi10:Aquifex_hydrogenase}. Other Aquificales, including its close relative \emph{H. thermophilus}, had previously been found to use 2-methylthiomenaquinone-7~\cite{Ishii87:quinone,Shima93:quinone,Stohr01:quinone}. Menaquinone (MK) has significantly lower redox potential than ubiquinone (UQ), and, based on distributions of these two quinone types both across the tree of life~\cite{Nitschke95:quinones,Schutz00:cytochrome} and within clades known to bridge the anaerobic-aerobic domains, UQ was suggested to have emerged with the rise of atmospheric oxygen~\cite{Schoepp09:Menaquinones}. Membrane-dissolved quinones exchange electrons directly with the fumarate/succinate redox couple, which is respectively an electron acceptor or donor depending on the direction of this reaction~\cite{Nitschke95:quinones,Iverson99:FR}. The possible emergence of the higher potential UQ with the rise of oxygen may thus have allowed the fumarate-succinate interconversion to reverse to the oxidative direction, further facilitating reversal of the TCA cycle as a whole. Generally, whereas reduced MK is easily oxidized in the presence of oxygen, disrupting electron flow into biosynthesis, reduced UQ is stable in the presence of oxygen~\cite{Schoepp09:Menaquinones}.  The redox potential of the de-methylated version of menaquinone, DMK, lies between that of MK and UQ, possibly reflecting the microaerophilic character of \emph{A. aoeolicus}~\cite{Infossi10:Aquifex_hydrogenase}.

\subsection*{Nucleotides}

Assignment of biosynthetic pathways to pyrimidines and purines was mostly unambiguous in \emph{A. aeolicus}. At the substrate level there is little major variation in the biosynthesis of these compounds~\cite{Zhang08:Purine}, and the genome of \emph{A. aeolicus} shows complete gene sets for their synthesis~\cite{Deckert98:Aquifex_genome}. There is some ambiguity in the interconversion between differently substituted purines and pyrimidines due to the well-known broad substrate affinity of many of the enzymes involved (e.g.~\cite{Heppel51:Nucleotidase,Berg54:Nucleosides}). We therefore did not significantly modify the conservative broad assignments made by SEED in this sub-network. Experimental studies would probably be needed to elucidate the fine scale activity/regulation of these reactions if it were deemed important for a highly quantitative model. We again suggest that the broad affinity of enzymes interconverting differently substituted nucleobases indicates that the lower mass flux of these reactions (for example compared to the rTCA cycle) significantly reduces the benefit relative to the cost of using multiple more-specific enzymes.

There are a few noteworthy details in the biosynthetic pathways of nucleotides. The initial synthesis of the IMP backbone involves several steps in which formyl groups are incorporated, which can proceed either through an ATP mediated addition of free formate, or through donation of the formyl group by N$^{10}$-Formyl THF~\cite{Zhang08:Purine}. Archaea that possess tetrahydromethanopterin (H$_4$MPT) rather than tetrahydrofolate (THF) as their C1 carrier use free formate in purine synthesis because H$_4$MPT is not a good donor of formyl groups~\cite{White97:Purine,Maden00:folates}. Most other organisms use THF to transfer formyl groups in purine synthesis~\cite{Zhang08:Purine}, while \emph{E. coli} was found to possess both mechanisms~\cite{Marolewski94:Purine}. \emph{A. aeolicus} follows these trends and uses THF as the formyl donor during purine synthesis.

Another minor variation in the synthesis of purines involves the carboxylation of aminoidazole ribunucleotide (AIR). Like other bacteria, \emph{A. aeolicus} uses a 2-step incorporation of HCO$_3^-$ involving ATP hydrolysis for this reaction. By contrast, higher organisms use a 1-step incorporation of CO$_2$ in an ATP free system, in which the enzyme is moreover often fused to the enzyme catalyzing the subsequent reaction~\cite{Zhang08:Purine}.  Similar to enzyme replacements we have seen in other metabolic sub-systems, this may represent another adaptation selected because it improves both thermodynamic efficiency and pathway throughput, in this case also shifting dependence from HCO$_3^-$ to CO$_2$ as a secondary effect.

Pyrimidine synthesis in \emph{A. aeolicus} again reflects the primitive nature of its metabolism. In the first reaction in this pathway carbamoyl phosphate is synthesized from glutamine, HCO$_3^-$, and ATP. Experimental studies showed that in \emph{A. aeolicus} this three-part reaction is catalyzed by a heterotrimer enzyme with relatively inefficient coupling between the subunits~\cite{Ahuja01:Aquif_CP}. By contrast, in \emph{E. coli} two of those subunits are fused together, resulting in a heterodimer enzyme, while in mammals all three subunits plus the enzymes for the subsequent reactions to carbamoyl-aspartate and dihydroorotate are fused together into one large single subunit enzyme~\cite{Ahuja04:Aquif_CP}.  Paralleling the suggested ancestry of ATP citrate lyase, the collection of observations about pyrimidine synthesis suggest that the heterotrimer carbamoyl-phosphate synthetase of \emph{A. aeolicus} is more closely related to the ancestral enzyme for this reaction~\cite{Ahuja01:Aquif_CP,Ahuja04:Aquif_CP}. However, while the cost of improving kinetics through gene fusion is lower than other cases of duplication and divergence, the lower mass flux density of pyrimidine synthesis relative to core carbon fixation also reduces the benefit of fusion. This may help explain why these genes are also not fused in many other bacteria. For \emph{A.  aeolicus} another reason that fusion did not take place may be that the heterotrimer structure may provide additional stability under hyperthermophilic conditions~\cite{Ahuja04:Aquif_CP}.

\subsection*{Cellular encapsulation}

The cellular encapsulation of \emph{A. aeolicus} consists of three main components: phospholipid membranes, a peptidoglycan cell wall, and lipopolysaccharide. \emph{A. aeolicus} has the full gene complement for the standard diaminopimelate-based variant of peptidoglycan synthesis that is common to gram-negative bacteria~\cite{Schleifer72:peptidoglycan}, but leaves substantial gaps within lipopolysaccharide synthesis pathways. These pathways remain an important area of experimental study, as they have required a large number of gap-fills for which we lacked overall context in the curation process. Of the three components of encapsulation, the composition of phospholipids contains the most information on the ecology of \emph{A. aeolicus}.

\subsubsection*{Lipid biosynthesis}

Lipid metabolism represents the single largest sub-system within the reconstructed network of \emph{A. aeolicus}, containing nearly 200 out of $\sim$760 reactions. As mentioned this is partly due to the fact that we explicitly represent each reaction within this sub-system, and partly due to the fact that \emph{A. aeolicus} has a complex and diverse lipid composition (see Supplementary Table I)~\cite{Jahnke01:Aquifex_FA}. However, the elongation of all fatty acid chains is a polymerization sequence in which 2-carbon units (from acetyl-CoA) are added, and then reduced, through a repeated sequence of the same 7 reactions catalyzed by the same 8 enzymes, with only the length of the fatty acid tail away from the reaction site varying~\cite{White05:FA_biosynth}. Much of the size of this sub-network thus reflects representation in the model rather than the associated genome content.

In addition, the fatty acid sub-network is further expanded due to the fact that \emph{A. aeolicus} uses fatty acid chains that contain methyl groups, propyl rings, and unsaturated bonds at different positions~\cite{Jahnke01:Aquifex_FA}. Each of these different substitutions is introduced at a different point during the elongation process, resulting in an expanding set of intermediates that is tracked in the network prior to the output of chains of different lengths and substitutions into the final lipid assembly process.

Finally, the lipid content of \emph{A. aeolicus} is unusual among bacteria for containing both phospho-ester and phospho-ether lipids~\cite{Jahnke01:Aquifex_FA}. In general, most bacteria use fatty acid-based ester lipids, while archaea use isoprenoid-based ether lipids~\cite{Lengeler99:BP}. The additional use of fatty acid ether lipids by \emph{A. aeolicus} thus represent a sort of intermediary strategy.

Altogether lipid biosynthesis can be thought of as a compact and highly modular system that distributes 2-carbon units over a set of states of different lengths and substitution patterns that can be varied depending on environmental context. Methyl group side-chains, unsaturated bonds or cyclopropane rings can be used to modify the fluidity of the membrane, while cyclopropane rings may also be used to adapt to lower pH~\cite{Zhang08:membrane_bacteria}. The linkage of fatty acids to the the glycerol backbone can in turn be varied between ether or ester bonds to modify the permeability of the membrane~\cite{Valentine07:archaea_energy}. Apart from basic inputs and final assembly (isoprenoids vs. fatty acids, ethers vs. esters), the regulation of lengths and substitution patterns appears to be the main factors permitting wide variability in lipid composition. The diverse and varied composition of \emph{A. aeolicus} lipids, including both ester and ether linkages, appears to reflect the ``stressed'' hyperthermophilic conditions of the hydrothermal vents and springs where it lives.

\subsection*{Energy metabolism}

Energy metabolism has the highest mass flux density of all cellular processes, because it generates the global energetic driving forces (both reductants and ATP) for all subsequent metabolic interconversion. The energy metabolism of \emph{A. aeolicus} represents one of its most studied aspects~\cite{Guiral12:Aquifex_rev}, and as we will demonstrate, the effects of kinetic optimization can be seen throughout.

An autotroph such as \emph{Aquifex} should be more sensitive to energy-metabolism optimization than more commonly-studied heterotrophic models, because unlike a heterotroph, which can obtain energy from organics and may use fermentation, autotrophs obtain all energy through purely respiratory interconversion of inorganics.  The free energy density available from inorganic redox couples may also be as much as an order of magnitude lower than that provided by sunlight used by photoautotrophs. Together these effects should create a significant selective advantage for improving the kinetics of energy metabolism in chemoautotrophs, by improving growth rate.

The apparent effects of improving kinetics can be seen at all levels of the energy metabolism of \emph{A. aeolicus}.  Many respiratory proteins are organized in polycistronic operons in the genome, and are subsequently assembled into super-complexes once functionally expressed. Moreover, whereas interactions between components of other known respiratory super-complexes are generally rather weak, in \emph{A. aeolicus} these super-complexes are found to be exceptionally stable~\cite{Peng03:Aquif_complexI,Guiral09:aquif_RespChains,Guiral12:Aquifex_rev}. While this is probably in part an adaptation to growth at high temperature, it also likely improves the overall kinetics of energy conversion sequences due to increased effective concentration of intermediates within each sequence.

\emph{A. aeolicus} has a versatile and diverse energy metabolism. Molecular hydrogen is sufficient as a sole electron donor, although it can in some cases be supplemented by hydrogen sulfide (H$_2$S).  A variety of compounds can act as terminal electron acceptors. Molecular oxygen (O$_2$) is the major electron acceptor, and under conditions so far tested it appears to be obligatory.  The metabolic network also indicates that nitrate (NO$_3^-$) can be used as electron acceptor, with the complete sequence for conversion to ammonia (NH$_3$) present~\cite{Deckert98:Aquifex_genome}. While this is in accordance with observations for other Aquificales~\cite{Vetriani04:Thermovibrio}, nitrate has so far not been reported as an electron acceptor for \emph{A. aeolicus}~\cite{Huber06:Aquificales,Guiral12:Aquifex_rev}.
 
An observation that H$_2$S cannot replace H$_2$ as sole electron donor may be explained by the existence of tightly-coupled respiratory super-complexes that prevent uptake of intermediates in the respiratory sequence~\cite{Guiral09:aquif_RespChains,Prunetti10:Aquifex}. Electrons are transfered from H$_2$ into metabolism at three main points, Hydrogenases I, II \& III, from where they enter the membrane quinone pool (Hydrogenase I, II) or are directly transferred to ferredoxin in the cytoplasm (soluble Hydrogenase III)~\cite{Guiral05:Aquifex_H_metab}. In the hydrogenase I respiratory chain, the quinones subsequently transfer the electrons to a cytochrome \emph{bc}$_1$ complex, which in turn reduces O$_2$ to water~\cite{Guiral03:Hydrogenase_Aquifex,Peng03:Aquif_complexI,Schutz03:Naphthoquinol}. In the hydrogenase II respiratory chain, the quinones instead transfer the electrons to the sulfur reductase complex, which in turn reduces elemental sulfur (and perhaps S$_4$O$_6^{2-}$) and produces H$_2$S~\cite{Nubel00:Aquifex_SGOR,Guiral05:aquif_Hcomplx}. H$_2$S can then subsequently be re-oxidized by a sulfide quinone reductase complex that transfers the electrons through quinones (and a cytochrome \emph{bc}$_1$ complex) into oxygen, which is reduced to water~\cite{Nubel00:Aquifex_SGOR,Guiral09:aquif_RespChains,Prunetti10:Aquifex}.
 
Sulfur has a dynamic and varied role in the energetics of \emph{A. aeolicus}, likely in part because of its ability to exist in a wide range of oxidation states~\cite{Wald62:periods}. In addition to acting as electron donor (H$_2$S, possibly S$^0$), several sulfur compounds are capable of acting as electron acceptors. Elemental sulfur (S$^0$) and tetrathionate (S$_4$O$_6^{2-}$) act as electron acceptors at the hydrogenase I complex~\cite{Guiral05:aquif_Hcomplx}. Thiosulfate (S$_2$O$_3^{2-}$) is in turn putatively oxidized by the Sox multi-enzyme system~\cite{Verte02:Sox,Ghosh09:SoxEvol,Guiral12:Aquifex_rev}, which has been described in another member of the Aquificaceae, \emph{H.  thermophilus}~\cite{Sano10:Sox}. The Sox system has also been described in other thermophiles that share with \emph{A. aeolicus} both the equivalent set of Sox genes in their genomes, and the characteristic of producing cytoplasmic sulfur globules under certain growth conditions~\cite{Miyake07:Thiosulfate}. Finally, \emph{A.  aeolicus} also possesses several rhodanese complexes possibly associated with cyanide detoxification~\cite{Giuliani07:rhodanese,Giuliani10:rhodanese}, as well as an ATP sulfurylase~\cite{Yu07:APS} possibly involved in sulfite oxidation~\cite{Guiral12:Aquifex_rev}. However, due to the complexity of sulfur chemistry, its roles in \emph{A. aeolicus} energy metabolism remain to be fully mapped out~\cite{Guiral12:Aquifex_rev}.

The energy metabolism of \emph{A. aeolicus} connects to its biosynthetic pathways mainly through the membrane quinone pool (with DMK-7 as the main component, see previous). Quinones reduce other biosynthetic reductant carriers such as nicotinamides (NAD, NADP), flavins (FAD), and ferredoxins, while in special cases also directly driving metabolic conversions (such as fumarate reduction to succinate).  Biosynthesis thus produces a net flux of electrons into the cell, which gives \emph{A. aeolicus} its reducing character. This electron influx is distributed across acceptors that range in character from purely anabolic (CO$_2$ during fixation and subsequent biosynthesis) to purely energetic (O$_2$), with some playing intermediary roles (some reduced nitrogen and sulfur is needed for biosynthesis). Shifts in character of these latter intermediary forms can be seen from the fact that \emph{A. aeolicus} growing on elemental sulfur will emit H$_2$S upon reaching the stationary phase of growth~\cite{Guiral05:aquif_Hcomplx}. Finally, global charge balance in the cell is maintained by combining this electron influx with a proton influx, which is captured by the ATP synthase to generate ATP~\cite{Peng06:Aquifex_ATP}.

\subsection*{Outlook and future directions}

We have used PMA to reconstruct the whole-genome metabolic network of \emph{A. aeolicus}, and have shown that it uses the likely ancestral pathways within many sub-systems. By reconstructing the evolutionary sequences among pathways, we were also able to analyze the evolutionary driving forces that shaped various sub-systems. We have highlighted throughout the way selection for improved kinetics and/or improved thermodynamic efficiency has shaped the network. Comparing different sub-networks reveals a tradeoff in the costs versus the benefits of these innovations, which apparently depends strongly on the relative mass flux density of sub-systems. Extending these analyses to other metabolic sub-systems and to the evolutionary history of other organisms will improve our understanding of how tradeoffs between performance gains and their associated costs generally contributed to fitness in the earliest stages of cellular life.

\section*{Acknowledgments}

Parts of this work were performed under support from NSF FIBR grant
nr.~0526747 -- The Emergence of Life: From Geochemistry to the Genetic
Code. RB was further supported by an SFI Omidyar Fellowship at the
Santa Fe Institute.  ES thanks Insight Venture Partners for support.

\bibliographystyle{unsrt}

\end{document}